\documentclass[amsmath,amssymb,twocolumn,pre,nofootinbib,floatfix]{revtex4}
\usepackage{bm}
\usepackage{color}   % needed when we import xfig .pstex files
\usepackage{epsfig}
\usepackage{subfigure}
\usepackage{graphicx}
\usepackage{wasysym} % for \ocircle
\usepackage{mathrsfs}
\usepackage{psfrag}
\psfrag{w100}{($1$,$0$,$0$)}
\psfrag{w010}{($0$,$1$,$0$)}
\psfrag{w0w0w1}{($w_1$=$0$,$w_2$=$0$,$w_3$=$1$)}
\psfrag{psi0}{$\bfpsi_0$}
\psfrag{psi1}{$\bfpsi_1$}
\psfrag{psi2}{$\bfpsi_2$}
\psfrag{psi3}{$\bfpsi_3$}
\psfrag{psi4}{$\bfpsi_4$}
\psfrag{psi5}{$\bfpsi_5$}
\psfrag{psi6}{$\bfpsi_6$}
\psfrag{psi7}{$\bfpsi_7$}
\psfrag{psiL0}{$\bfpsi^L_0$}
\psfrag{psiL1}{$\bfpsi^L_1$}
\psfrag{psiL2}{$\bfpsi^L_2$}
\psfrag{psiL3}{$\bfpsi^L_3$}
\psfrag{psiL4}{$\bfpsi^L_4$}
\psfrag{psiL5}{$\bfpsi^L_5$}
\psfrag{psiL6}{$\bfpsi^L_6$}
\psfrag{psiL7}{$\bfpsi^L_7$}
\psfrag{gamma0}{$\gamma_0$}
\psfrag{gamma0-gamma2}{$\gamma_0$-$\gamma_2$}
\psfrag{gamma0-gamma5}{$\gamma_0$-$\gamma_5$}
\psfrag{gamma1}{$\gamma_1$}
\psfrag{gamma1-gamma2}{$\gamma_1$-$\gamma_2$}
\psfrag{gamma2}{$\gamma_2$}
\psfrag{gamma2-gamma3}{$\gamma_2$-$\gamma_3$}
\psfrag{gamma3}{$\gamma_3$}
\psfrag{gamma4}{$\gamma_4$}
\psfrag{gamma5}{$\gamma_5$}
\psfrag{gamma6}{$\gamma_6$}
\psfrag{gamma7}{$\gamma_7$}
\psfrag{gamma0}{$\gamma_0$}
\psfrag{g0}{$\gamma_0$}
\psfrag{g0-g2}{$\gamma_0$-$\gamma_2$}
\psfrag{g0-g5}{$\gamma_0$-$\gamma_5$}
\psfrag{g1}{$\gamma_1$}
\psfrag{g1-g2}{$\gamma_1$-$\gamma_2$}
\psfrag{g2}{$\gamma_2$}
\psfrag{g2-g3}{$\gamma_2$-$\gamma_3$}
\psfrag{g3}{$\gamma_3$}
\psfrag{g4}{$\gamma_4$}
\psfrag{g5}{$\gamma_5$}
\psfrag{g6}{$\gamma_6$}
\psfrag{g7}{$\gamma_7$}
\psfrag{w0}{$\bfw_0$}
\psfrag{w1}{$\bfw_1$}
\psfrag{w2}{$\bfw_2$}
\psfrag{w3}{$\bfw_3$}
\psfrag{w4}{$\bfw_4$}
\psfrag{w5}{$\bfw_5$}
\psfrag{w6}{$\bfw_6$}
\psfrag{w7}{$\bfw_7$}
\psfrag{Ma0}{\large $X_0$}
\psfrag{Ma1}{\large $X_1$}
\psfrag{M11}{\large $M_{1 1}$}
\psfrag{M10}{\large $M_{1 0}$}
\psfrag{w001}{($0$,$0$,$1$)}
\psfrag{c1}{$1$}
\psfrag{c2}{$2$}
\psfrag{c3}{$3$}
\psfrag{c4}{$4$}
\psfrag{c5}{$5$}
\psfrag{c6}{$6$}

\usepackage{afterpage}
\usepackage{float}
\usepackage{graphics}

\usepackage{pstricks}
\usepackage{pst-node}
\graphicspath{{./figs}}
\usepackage{array}

\newcommand{\fig}[1]{Fig.\ \ref{fig:#1}}
\newcommand{\figs}[1]{Figs.\ \ref{fig:#1}}

\newcommand{\eqn}[1]{Eq.\ (\ref{Eqn:#1})}
\newcommand{\eqns}[1]{Eqs.\ (\ref{Eqn:#1})}
\newcommand{\eqnref}[1]{(\ref{Eqn:#1})}

\newcommand{\bfone}{{\bm 1}}

\newcommand{\bfpsi}{\bm{\psi}}
\newcommand{\bfpi}{\bm{\pi}}

\newcommand{\bfp}{\bm{p}}
\newcommand{\bfw}{\bm{w}}
\newcommand{\bfx}{\bm{x}}
\newcommand{\bfy}{\bm{y}}

\newcommand{\bfl}{\bm{1}}

\newcommand{\operp} {{{\scriptstyle \bigcirc} \!\!\!\! {\scriptscriptstyle \perp}}}
\newcommand{\com}[1]{}

\newcommand{\myxx}[1]{}
\newcommand{\bvec}{\overrightarrow}

\makeatletter

\newdimen\psparallelogramsep
\def\psset@parallelogramsep#1{\pssetlength\psparallelogramsep{#1}}
\psset@parallelogramsep{3mm}

\def\psparallelogrambox{\pst@object{psparallelogrambox}}

\def\psparallelogrambox@i{\pst@makebox\psparallelogrambox@ii}

\def\psparallelogrambox@ii{%
\begingroup
\pst@useboxpar
\pst@dima=\pslinewidth
\advance\pst@dima by \psframesep
\pst@dimc=\wd\pst@hbox\advance\pst@dimc by \pst@dima
\pst@dimb=\dp\pst@hbox\advance\pst@dimb by \pst@dima
\pst@dimd=\ht\pst@hbox\advance\pst@dimd by \pst@dima
\setbox\pst@hbox=\hbox{%
\ifpsboxsep\kern\pst@dima\fi
\begin@ClosedObj
\addto@pscode{%
\psk@cornersize
\pst@number\pst@dima neg
\pst@number\pst@dimb neg
\pst@number\pst@dimc
\pst@number\pst@dimd
.5
\pst@number\psparallelogramsep
\tx@Parallelogram}%
\def\pst@linetype{2}%
\showpointsfalse
\end@ClosedObj
\box\pst@hbox
\ifpsboxsep\kern\pst@dima\fi}%
\ifpsboxsep\dp\pst@hbox=\pst@dimb\ht\pst@hbox=\pst@dimd\fi
\leavevmode\box\pst@hbox
\endgroup}

\newdimen\autotop  \newdimen\autobottom    \newdimen\autosize

\def\autorightleft#1#2{%
\setbox0=\hbox{#1}\autotop=\wd0\setbox0=\hbox{#2}\autobottom=\wd0%
\ifdim\autobottom>\autotop\autosize=\autobottom\else\autosize=\autotop\fi%
\advance\autosize by 0.2em%
\mathop{\vcenter{\hbox{\ooalign{\raise1ex%
          \hbox{$\hbox to \autosize{\hspace*{.05em}\rightarrowfill}$}\crcr%
                $\hbox to \autosize{\leftarrowfill\hspace*{.05em}}$}}}}%
            \limits^{\hbox{#1}}_{\hbox{#2}}}

\def\rightleft#1#2{%
\setbox0=\hbox{#1}\autotop=\wd0\setbox0=\hbox{#2}\autobottom=\wd0%
\ifdim\autobottom>\autotop\autosize=\autobottom\else\autosize=\autotop\fi%
\advance\autosize by 2em%
\mathop{\vcenter{\hbox{\ooalign{\raise1ex%
          \hbox{{\hspace*{.05em}$\longrightarrow$}}\crcr%
                {$\longleftarrow$\hspace*{.05em}}}}}}%
            \limits^{\hbox{#1}}_{\hbox{#2}}}

\pst@def{Parallelogram}<{%
/ParallelogramA {
x1 pgs sub y1 moveto
x1 y2 lineto
x2 pgs add y2 lineto
x2 y1 lineto
x1 pgs sub y1 lineto
closepath} def
/pgs ED
CLW mul
/a ED
3 -1 roll
2 copy gt { exch } if
a sub
/y2 ED
a add
/y1 ED
2 copy gt { exch } if
a sub
/x2 ED
a add
/x1 ED
1 index 0 eq {pop pop ParallelogramA } { OvalFrame } ifelse}>

\makeatletter

\begin{document}

\title{Efficient uncertainty minimization for fuzzy spectral clustering}
\author{Brian S. White}
\email{bsw27@cornell.edu}
\author{David Shalloway}
\email{dis2@cornell.edu}
\affiliation{Biophysics Program, Department of Molecular Biology and Genetics, Cornell University, Ithaca, New York 14853}

%%% ChangeLog

%%% arXiv v2
%%% - New implementation of and motivation for initial solver
%%% - Referenced PCCA+
%%% - Removed sections on computational implementation and
%%%   treatment of asymmetric transition matrices (due to space)
%%% - General tidying up

%%% arXiv v5
%%% - Changed title to emphasize fuzzy and downplay MDC
%%% - Changed abstract to emphasize fuzzy spectral as opposed to
%%%   spectral vs traditional.  Also mention that uncertainty
%%%   minimization has been generalized to apply to any Gamma
%%%   satisfying dynamical interpretation.
%%% - Changes to results section
%%%     - Removed results on FCPS data sets
%%%       i.e., removed old Fig 3 and Table I
%%%     - Added fuzzy results for different gammas, including
%%%       a symmetric gamma with an exponential similarity measure and
%%%       an asymmetric normlzied gamma with an exponential similarity.
%%%       This replaces Fig 3, but has become Fig 4.
%%% - Removed subsection on recursion, which has nothing to do with
%%%   uncertainty minimization and is fairly intuitive.
%%% - Removed from superfluous computational detail regarding shift-
%%%   and-invert spectral transformation.

\begin{abstract}
Spectral clustering uses the global information embedded in eigenvectors of an inter-item similarity matrix to correctly identify clusters of irregular shape, an ability lacking in commonly used approaches such as $k$-means and agglomerative clustering.  However, traditional spectral clustering partitions items into hard clusters, and the ability to instead generate fuzzy item assignments would be advantageous for the growing class of domains in which cluster overlap and uncertainty are important. Korenblum and Shalloway [Phys. Rev. E {\bf 67}, 056704 (2003)] extended spectral clustering to fuzzy clustering by introducing the principle of uncertainty minimization.  However, this posed a challenging non-convex global optimization problem that they solved by a brute-force technique unlikely to scale to data sets having more than $O(10^2)$ items. Here we develop a new method for solving the minimization problem, which can handle data sets at least two orders of magnitude larger.  In doing so, we elucidate the underlying structure of uncertainty minimization using multiple geometric representations. This enables us to show how fuzzy spectral clustering using uncertainty minimization is related to and generalizes clustering motivated by perturbative analysis of almost-block-diagonal matrices. Uncertainty minimization can be applied to a wide variety of existing hard spectral clustering approaches, thus transforming them to fuzzy methods.
\end{abstract}

\pacs{02.70.-c, 02.70.Hm, 02.50.Fz, 89.75.Kd}
\maketitle

\renewcommand{\baselinestretch}{1.2}

\section{Introduction}
\label{section:introduction}

Coarse-graining data {\em items} $i$ ($1 \le i \le N$) into \emph{clusters} $\alpha$ ($1 \le \alpha \le m$) is important for large-scale data analysis~\cite{Jain:99, Everitt:01, Xu:05}. For example, clustering genes according to their microarray expression profiles allows biologists to subsequently infer potential {\em cis}-regulatory elements from sequence commonalities within the clusters~\cite{Cho:98}.  Clustering typically proceeds from a symmetric $N \times N$ \emph{similarity matrix} $S$, where the non-negative off-diagonal element $S_{ij}$ provides an inverse indicator of the ``distance'' $d_{ij}$ between items $i$ and $j$. The primary input (e.g., the alignment scores from sequence comparisons or edge weights of a graph) may directly define the $S_{ij}$.  Alternatively, the data may consist of $N_D$ \emph{properties} for each item that can be embedded in a dataspace. For example, in microarray analysis each gene is an item, and its
properties are its $N_D$ expression levels under $N_D$ different conditions. In that case, the $d_{ij}$ are derived from the (not-necessarily Euclidean) distances between the items in the dataspace.

\emph{Spectral clustering} methods (\cite{Spielman:96, vonLuxburg:07} for history and review) analyze the eigensystem of a \emph{transition} (or
\emph{Laplacian}) \emph{matrix} $\Gamma$, which is derived from $S$.
 Since the eigensystem depends globally on the entire data set, spectral methods have a perspective lacking in commonly used methods
 such as $k$-means and agglomerative clustering~\cite{Everitt:01}, which directly analyze the $S_{ij}$.  Their dependence on pairwise similarities leads them to impose characteristic cluster shapes; e.g., $k$-means and complete-linkage clustering generate convex clusters while single-linkage clustering generates unbalanced and straggly clusters~\cite{Everitt:01}. These shapes may not reflect the true geometries of the problem, such as the irregular boundaries of a subject within an image~\cite{Shi:97}. The ability of spectral methods to generate arbitrary cluster shapes lets them outperform $k$-means across several benchmarks~\cite{Kamvar:03,Ng:02,Korenblum:03}. And as we will see, they can also determine the optimal number of clusters automatically.

$\Gamma$ typically satisfies~\cite{ft:01,ft:02}
\begin{subequations}
\label{Eqn:dynamical_properties}
\begin{eqnarray}
\Gamma & = & \Gamma^S \cdot D^{-1}_\pi \label{Eqn:gamma_form} \\
\Gamma^S_{ij} & = & -S_{ij} \qquad (i \ne j) \\
\Gamma^S_{ii} & = & \sum_{j \ne i} S_{ji} \label{Eqn:diagonal_S}\\
\bfl \cdot \Gamma & = & 0 \label{Eqn:probability_conservation} \;,
\end{eqnarray}
\end{subequations}
where $D_\pi$ is a diagonal normalizing matrix with non-negative elements satisfying $\operatorname{Tr}(D_\pi)=1$, $\bfl$ is the item-space vector having all components equal to one, and $\cdot$ denotes the normalized item-space inner product:
\[
\bfx \cdot {\bfy} \equiv N^{-1} \sum_{i=1}^{N} x_i y_i \;.
\]
These conditions emerge when spectral clustering methods are used to approximate ``min-cut'' graph partitioning solutions~\cite{Donath:73, Fiedler:73} or when they are motivated by discrete-~\cite{Meila:00, Meila:01,Belkin:03, Weber:04, Nadler:06} or continuous-time~\cite{Korenblum:03} dynamical models. [The first two motivations lead to analysis of the Markov matrix $T \equiv I-\Gamma$ (where $I$ is the identity matrix), which satisfies $\bfl \cdot T= \bfl$ rather than \protect\eqn{probability_conservation}. But since the eigenvectors of $T$ and $\Gamma$ are identical and the eigenvalues are simply related, the same analysis applies with inconsequential changes.]

\eqns{dynamical_properties} imply
\begin{subequations}
\label{Eqn:spectral_properties}
\begin{eqnarray}
\gamma_0 & = & \, 0\\
\bfpsi^R_0 & \equiv & N \bfpi     \label{Eqn:right_0_eigenvector} \\
\bfpsi^L_0 & = & \bfl \label{Eqn:left_0_eigenvector}\\
\bfpsi^L_n & =& D_\pi^{-1} \cdot \bfpsi^R_n  \; ,  \label{Eqn:left_n_eigenvector}
\end{eqnarray}
\end{subequations}
where $\bfpsi^L_n$ and $\bfpsi^R_n$ are the bi-orthogonal left and right eigenvectors of $\Gamma$, which we normalize such that $\bfpsi^L_m \cdot \bfpsi^R_n = \delta_{mn}$ and $\bfpsi^L_n \cdot \bfpsi^L_n =1$, and $\bfpi$ is the right equilibrium probability vector satisfying $\sum_i \pi_i =1$. It follows that $(D_\pi^{-1})_{ii}=\pi_i^{-1}$. \eqns{dynamical_properties} also imply that $\Gamma_{ij} \pi_j = \Gamma_{ji} \pi_i$ (i.e., that detailed balance holds), which ensures the reality and non-negativity of the eigenvalues~\cite{ft:03}.

Spectral methods begin by embedding each item $i$ into the \emph{low-frequency} (or \emph{clustering}) \emph{subspace} ${\mathbb R}^m$ using as coordinates the $m$ low-frequency vector components of $\bvec{\psi}^L(i) \equiv [\psi^L_0(i), \, \psi^L_1(i) \, \ldots \, \psi^L_{m-1}(i)]$~\cite{ft:04}. These are then used to identify $m$ clusters~\cite{ft:05}. Clustering (i.e., spatial coarse-graining) is possible only if there is a gap in the distribution of the similarities $S_{ij}$~\cite{ft:06}.

The dynamical interpretation of spectral clustering provides a way to find a gap if it exists: Each cluster is viewed as a metastable state of a diffusive relaxation process governed by $\Gamma$~\cite{ft:07}%
\begin{equation}
\label{Eqn:microscopic_dynamics}
\frac{d \bfp(t)}{dt} = - \Gamma \cdot \bfp(t)\;,
\end{equation}
where $\bfp(t)$ is a time-dependent probability vector over the discrete space of items [i.e., $p_i(t)$ is the probability of occupation of item $i$ at time $t$], $-\Gamma_{ij}$ is the stochastic transition rate from item $j$ to $i$, and \eqns{diagonal_S} and \eqnref{probability_conservation} ensure that probability is conserved. Because of the inverse relationship between eigenvector ``wavelength'' and eigenvalue, a spatial-scale gap in the distribution of the $S_{ij}$ will appear as a time-scale \emph{spectral gap}:
\begin{equation}
\label{Eqn:spectral_gap}
 0 = \gamma_0 < \gamma_1 < ... < \gamma_{m-1} \ll \gamma_m \;.
\end{equation}
The gap between $\gamma_{m-1}$ and $\gamma_m$ indicates the existence of $m$ clusters. When a spectral gap exists, the long-wavelength, clustering eigenvectors $\bfpsi^L_{n<m}$ will contain the information needed for clustering~\cite{ft:08}.
%com
\begin{center}
  \begin{figure*}
    \begin{center}
      \begin{math}
        \begin{array}{ccc}
          \multicolumn{3}{c}{
            \subfigure[]{
              \label{fig:spiral_dataset}
              \includegraphics{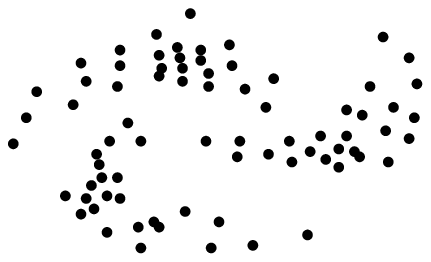}
            }
          }
          \\
          \subfigure[]{
            \label{fig:spiral_eig0}
            \scalebox{1.0}{
              \includegraphics{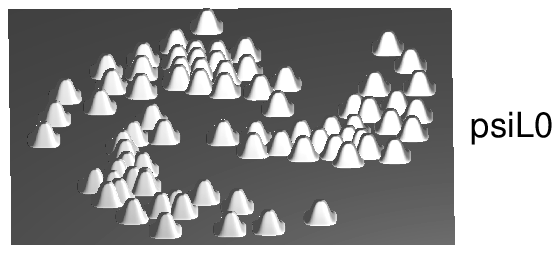}
            }
          }
          &
          \subfigure[]{
            \label{fig:spiral_eigs}
            \scalebox{1.0}{
              \includegraphics{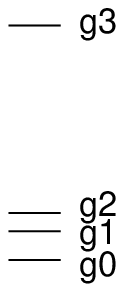}
            }
          }
          &
          \subfigure[]{
            \label{fig:spiral_eig1}
            \scalebox{1.0}{
              \includegraphics{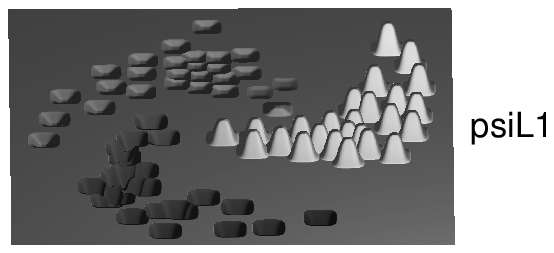}
            }
          }
          \\
          \subfigure[]{
            \label{fig:spiral_eig2}
            \scalebox{1.0}{
              \includegraphics{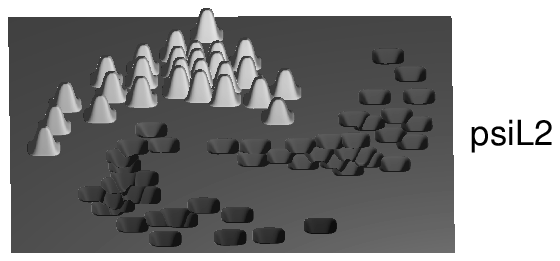}
            }
          }
          &
          &
          \subfigure[]{
            \label{fig:spiral_eig3}
            \scalebox{1.0}{
           \includegraphics{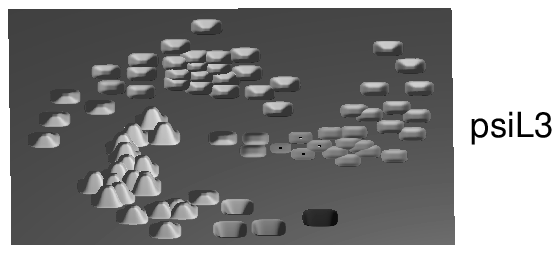}
            }
          }
        \end{array}
      \end{math}
\caption{The ``spiral'' clustering problem and its eigensystem.  (a) The two-dimensional embedding of the spiral data set in dataspace.  (b), (d), (e), (f) The (positive or negative) heights of the cones indicate the values of the clustering eigenvectors $\bfpsi^L_0$, $\bfpsi^L_1$, and $\bfpsi^L_2$, and of the first non-clustering eigenvector, $\bfpsi^L_3$ of the data set's $\Gamma$.  (c) The corresponding eigenvalues. Unless otherwise noted, figures are based on the $\Gamma$ matrix defined by Korenblum and Shalloway~\protect\cite{Korenblum:03} [see \protect\eqns{Gamma_KS}].}
    \label{fig:spiral_eigensystem}
    \end{center}
  \end{figure*}
\end{center}

For example, \fig{spiral_eigensystem} illustrates the $m=3$ ``spiral'' clustering problem posed by 77 items embedded in a two-dimensional dataspace and the corresponding eigensystem of the $\Gamma$ matrix of Ref.\ \cite{Korenblum:03} [see \eqns{Gamma_KS} below]. Panel (a) shows the spatial locations of the items, and it is subjectively evident that there are three interlocking clusters. Correspondingly, as predicted by \eqn{spectral_gap}, there is a gap between $\gamma_2$ and $\gamma_3$ [panel (c)]. The clustering eigenvectors, $\bfpsi^L_1$ [panel (d)] and $\bfpsi^L_2$ [panel (e)], vary significantly only at the cluster boundaries and follow their distorted shapes.  Thus, the shapes of the clusters defined using these eigenvectors will not be artificially restricted. In contrast, the non-clustering eigenvectors such as $\bfpsi^L_3$ [panel (f)] have large variations within clusters and thus are not used in the clustering analysis.

It remains to define the clustering from the clustering eigenvectors.  Hard spectral clustering approaches do so simply by applying non-spectral methods such as $k$-means within the clustering subspace~\cite{Ng:02}. However, there are problems where hard partitioning is neither necessary nor ideal, for example, the separation of cell subpopulations by fluorescence activated cell sorting (FACS)~\cite{Jeffries:08}, automated biological database curation~\cite{Paccanaro:06}, complex network analysis~\cite{Reichardt:04}, and gene expression analysis~\cite{Gasch:02}.  Such problems require {\em fuzzy clustering} that can represent uncertainty and overlapping clusters.

Non-spectral fuzzy clustering methods have already been applied to such problems~\cite{Reichardt:04,Gasch:02}, but spectral fuzzy methods could be advantageous because of their added ability to cope with irregular cluster boundaries (such as those within FACS dataspaces~\cite{Jeffries:08}). Moreover, fuzziness could provide further benefit even in areas where hard spectral clustering has already been applied.  For example,  Paccanaro et al.~\cite{Paccanaro:06} have used hard spectral clustering to faithfully reproduce many of the superfamily classifications from a subset of the SCOP protein database~\cite{Murzin:95}; a fuzzy spectral approach would add the ability to assess the certainty of such classifications.

Formally, fuzzy clusterings are described by {\em assignment vectors} $\bfw_\alpha \equiv [w_\alpha (1), w_\alpha(2), \ldots, w_\alpha(N)]$, where $w_\alpha(i)$ is the probability that item $i$ is a member of cluster $\alpha$, and therefore must satisfy the probabilistic constraints
\begin{subequations}
\label{Eqn:w_constraints}
\begin{eqnarray}
w_\alpha(i) & \ge & 0 \qquad (\forall \;\alpha, i)
\label{Eqn:discrete_w_inequality_constraints}  \\
\sum_{\alpha} w_\alpha(i) &=& 1 \qquad (\forall \;i) \;.
\label{Eqn:discrete_w_equality_constraints}
\end{eqnarray}
\end{subequations}
To define these in a spectral context, following Ref.~\cite{Korenblum:03} we use the low-frequency clustering eigenvectors as a linear basis for the $\bfw_\alpha$~\cite{ft:09}:
\begin{equation}
\label{Eqn:w_psi_discrete}
\bfw_\alpha = \sum_{n=0}^{m-1} M_{\alpha n} \, \bfpsi^L_n \equiv \bvec{M}_\alpha \circ \bvec{\bfpsi^L} \;,
\end{equation}
where the $\bvec{M}_\alpha \equiv [M_{\alpha 0}, \,M_{\alpha 1},\, \ldots \, M_{\alpha (m-1)}]$ are $m$-vectors, $\bvec{\bfpsi^L} \equiv [\bfpsi^L_0, \bfpsi^L_1, \ldots,\bfpsi^L_{m-1}]$, and $\circ$ denotes the inner product over the low-frequency subspace:
\[
\bvec{x} \circ \bvec{y} = \sum_{n=0}^{m-1} x_n \, y_n \;.
\]

\eqn{w_psi_discrete} transforms the clustering problem to that of finding the ``best'' $\bvec{M}_\alpha$ subject to \eqns{w_constraints}.
Korenblum and Shalloway~\cite{Korenblum:03} proposed that this was the one that minimized overlap between assignment vectors:  Since the $\bfw_\alpha$ are non-negative and composed of only the long-wavelength $\bfpsi^L_{n<m}$, they will inevitably overlap each other and thus will give \emph{uncertain} (i.e., fuzzy) item-to-cluster assignments.  This uncertainty is minimized when the clusters' self-overlap is maximized.  The self-overlap (of cluster $\alpha$) can be quantified by the \emph{fractional cluster certainty} ${\overline\Upsilon}_\alpha(M) \;(1 \le \alpha \le m)$~\cite{Korenblum:03},
\begin{equation}
\label{Eqn:upsilon}
{\overline\Upsilon}_\alpha(M) \equiv  \frac{\langle \bfw_\alpha | \bfw_\alpha \rangle}{\langle \bfl | \bfw_\alpha \rangle} \qquad (N^{-1} \le {\overline\Upsilon}_\alpha(M) \le 1) \; ,
\end{equation}
where $M$ represents the components of all the $\bvec{M}_\alpha$ and bra-ket notation denotes the equilibrium-weighted inner product~\cite{ft:10}%
\begin{equation}
\label{Eqn:D_pi}
\langle \bfx | {\bfy} \rangle \equiv \bfx \cdot D_{\bfpi} \cdot \bfy \,.
\end{equation}
${\overline\Upsilon}_\alpha(M)=1$ when the cluster $\alpha$ is completely certain, i.e., $w_\alpha(i) = 0 \mbox{ or } 1$; the total certainty  is the product of the ${\overline\Upsilon}_\alpha(M)$ for all the clusters. Thus, the optimal $M$ is determined by \emph{uncertainty minimization} of the \emph{overall uncertainty objective function},
\begin{equation}
\label{Eqn:minimum_uncertainty_A}
 \Phi(M) \equiv - \sum_\alpha \log {\overline \Upsilon}_\alpha(M) \;,
\end{equation}
subject to the constraints of \eqns{w_constraints}. Korenblum and Shalloway showed that this procedure provided good fuzzy clusterings of a number of difficult problems. However, they solved the resulting challenging constrained, non-convex uncertainty minimization problem using a ``brute-force'' solver whose $O(m^2N^{m+1})$ computational complexity limited its application to modest-sized problems ($N=200$) and precluded application to the larger problems [e.g., $N \sim O(10^4)$] that emerge in areas such as gene microarray analysis~\cite{Eisen:98}.

A closely related approach was independently developed by Weber et al.~\cite{Weber:04}. They also used \eqn{w_psi_discrete}, but, instead of using uncertainty minimization, determined the $M$ through an efficient, but approximate,  method motivated by perturbative analysis of almost-block-diagonal matrices~\cite{Stewart:84}.  Their Perron Cluster Cluster Analysis (PCCA) defined the $\bfw_\alpha$ as ``membership functions'' that only approximate the probabilistic constraints of \eqns{w_constraints}. In PCCA the $M$ are determined algorithmically rather than by objective function optimization, and clusterings for different values of $m$ are accepted if the resultant approximation is regarded (by subjective criteria) to be adequate.  While approximate, this method had the advantage of being computationally simpler than the initial uncertainty minimization algorithm of Korenblum and Shalloway~\cite{Korenblum:03}.

Thus until now, \emph{practical}, \emph{exact} fuzzy spectral data clustering has remained elusive.
To resolve this problem, here we develop an efficient method for uncertainty minimization and show that it is  generally applicable to any spectral clustering method satisfying \eqns{dynamical_properties}, including popular asymmetric approaches based on random walks over graphs~\cite{Meila:00, Meila:01, Belkin:03, Weber:04, Nadler:06}.  Thus, we imbue a wide range of hard spectral clustering methods with the ability to represent fuzzy cluster assignments and, thereby, uncertainty and cluster overlap. In the process, we show that there are multiple geometric interpretations of the uncertainty minimization problem that can be used to illuminate its structure. Through these we relate uncertainty minimization to PCCA and extend the previously reported conditions under which the PCCA approximation is applicable.

\section{Computational Theory}
\label{section:computational}

Minimization of $\Phi(M)$ subject to the constraints of \eqns{w_constraints} poses a global, non-linear optimization problem in the $m^2$ degrees of freedom of $M$. To solve this it is convenient to reexpress
\eqn{minimum_uncertainty_A} explicitly in terms of the $\bvec{M}_\alpha$ as
\begin{equation}
\label{Eqn:minimum_uncertainty}
 \Phi(M) \equiv - \sum_\alpha \log {\overline \Upsilon}_\alpha(M)
= - \sum_\alpha \log \frac{\bvec{M}_\alpha \circ \bvec{M}_\alpha}{\bvec{M}_\alpha \circ \hat{\varepsilon}_0}
\;,
\end{equation}
where $\hat{\varepsilon}_0$ is the $m$-vector $(1,0, \ldots, 0)$, and we have used $\langle \bfw_\alpha | \bfw_\alpha \rangle = \bvec{M}_\alpha \circ \bvec{M}_\alpha$ and $\langle \bfl | \bfw_\alpha \rangle = \bvec{M}_\alpha \circ \hat{\varepsilon}_0$,  which follow from \eqns{left_0_eigenvector}, \eqnref{left_n_eigenvector}, and \eqnref{w_psi_discrete} and the bi-orthogonality of the eigenvectors. Similarly, we reexpress \eqns{w_constraints} in terms of the $\bvec{M}_\alpha$:
\begin{subequations}
\label{Eqn:M_constraints}
\begin{eqnarray}
w_\alpha(i) & = & \bvec{M}_\alpha \circ \bvec{\psi^L}(i) \ge 0 \qquad (\forall \;\alpha, i) \label{Eqn:discrete_inequality_constraints} \\
\sum_\alpha \bvec{M}_\alpha & = & \hat{\varepsilon}_0 \;.
\label{Eqn:discrete_equality_constraints}
\end{eqnarray}
\end{subequations}
Because $\Phi(M)$ is invariant under permutations of the indices associated with the clusters, its global minimum will have an $m!$-fold permutation degeneracy.

We now describe two geometric representations that illuminate the problem (Sec.~\ref{section:geometric}) and then show how to solve it in three steps:  (1) precondition $\Gamma$ to avoid numerical noise that can obfuscate spectral gaps when low-lying eigenvalues are nearly degenerate, to improve numerical efficiency, and to remove outliers (Appendix~\ref{appendix:precondition}), (2) find a zeroth-order solution  (Sec.~\ref{section:zeroth_order_solver}), and (3) iteratively refine using linear programming with a subset of the inequality constraints to determine the solution to the desired accuracy (Sec.~\ref{section:constrained_solver}).  Since the procedure explicitly uses only the $\bfpsi^L_n$, for notational convenience we subsequently denote them simply as the $\bfpsi_n$.

\subsection{Geometric representations of uncertainty minimization}
\label{section:geometric}
\subsubsection{Symmetric $M$-representation}
Each $\bvec{M}_\alpha$ may be regarded as the coordinates of a particle $\alpha$ in ${\mathbb R}^m$ with axes labeled $X_0, X_1, \ldots, X_{(m-1)}$.  \eqn{discrete_inequality_constraints} implies that the same $N$ inequality constraints act on each particle; thus they restrict each one to the same half-space in ${\mathbb R}^m$ bounded by a hypersurface passing through the origin and normal to $\bvec{\psi}(i)$.  The intersection of these half-spaces determines the feasible region as a convex polyhedral cone in the upper half of ${\mathbb R}^m$. Only a subset of the inequality constraints will actually bound the feasible region, since their satisfaction will automatically guarantee satisfaction of the other constraints.  And, as proved in Appendix
\ref{appendix:inequality_constraint_proof}, each particle lies on an edge of the polyhedral cone (i.e., is constrained by $m-1$ \emph{active} inequality constraints) at every local minimizer of $\Phi(M)$.

\begin{figure}[htpb]
\begin{math}
\begin{tabular}{cc}
\subfigure[]{\label{fig:cres_mspace}\scalebox{0.70}{
    \includegraphics{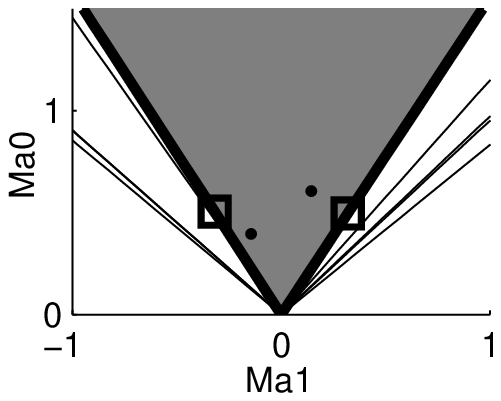}
}}
 &
\subfigure[]{\label{fig:cres_mspace_constrained}\scalebox{0.70} {
    \includegraphics{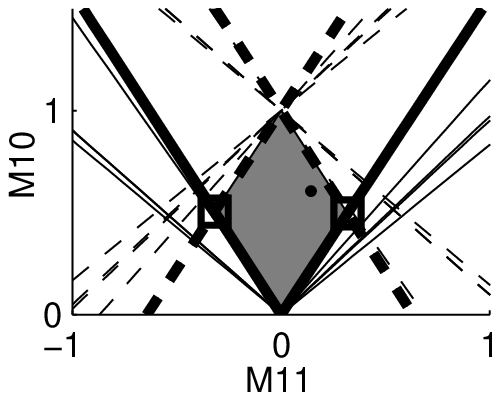}
}}
\end{tabular}
\end{math}
\caption{
     Symmetric and asymmetric $M$-representations of the $m=2$ ``crescentric'' problem~\cite{Korenblum:03,Everitt:01}. (a) \emph{Symmetric $M$-representation:} The diagonal lines indicate the boundaries formed by the inequality constraints. The two bold lines forming the narrowest cone (shaded) define the feasible region  in ${\mathbb R}^m= {\mathbb R}^2$. $\protect\bvec{M}_1$ and $\protect\bvec{M}_2$ are represented by dots. They are not independent since they are further constrained by the equality constraints of \eqn{discrete_equality_constraints}. The global minimum of the uncertainty objective function $\Phi(M)$ corresponds to the dots being located at the positions indicated by small squares, and the invariance under particle exchange corresponds to the permutation degeneracy discussed in the text. (b) \emph{Asymmetric $M$-representation:} The solid lines indicate the boundaries of the homogeneous inequality constraints acting on the free particle $\protect\bvec{M}^{\rm free}=\protect\bvec{M}_1$.  The dashed lines indicate the boundaries of the inhomogeneous inequality constraints that derive from the slave particle $\protect\bvec{M}_2$. Two of these (bold-dashed) lines cap the cone formed by the relevant homogeneous constraint boundaries (bold) to define a closed feasible polytope in ${\mathbb R}^{m(m-1)}={\mathbb R}^2$.  In this representation the single dot represents all $m(m-1)=2$ components of $\protect\bvec{M}^{\rm free}$. $\Phi(M)$ is minimized at either of the two permutation-degenerate solutions (small squares).}
     \label{fig:m_2}
\end{figure}

An example of this \emph{symmetric $M$-representation} for an $m=2$  problem (based on the ``crescentric'' bivariate data set of Ref.\ \cite{Everitt:01}) is shown in \fig{cres_mspace}. (It is only in the $m=2$ case that a simple graphical representation is possible; nonetheless it is useful for illustrating structural properties that also hold when $m>2$.)
In this case, the feasible region is bounded by only two lines corresponding to $\bvec{X} \circ \bvec{\psi}(i_<) = 0$ and $\bvec{X} \circ \bvec{\psi}(i_>) = 0$, where $i_<$ and $i_>$ are the minimizer and maximizer of $\psi_1(i)$, respectively. The global minimum of $\Phi$ corresponds to the unique (up to the permutation degeneracy) situation where each particle lies on the feasible region boundary while the equality constraints of \eqn{discrete_equality_constraints} are simultaneously satisfied. In \fig{cres_mspace}, this is when the points are located at the two squares on the boundary. The two ways of associating the particles with the squares corresponds to the 2-fold permutation degeneracy of the solution.
%
%\begin{center}
%\end{center}

\subsubsection{Asymmetric $M$-representation}
The $m$ particles in the symmetric $M$-representation are not independent because of the equality constraints [\eqn{discrete_equality_constraints}]. We  use these in the \emph{asymmetric $M$-representation} to explicitly eliminate the degrees of freedom of one {\em slave} particle that, without loss of generality, we take to be $\bvec{M}_m$:
\begin{equation}
\label{Eqn:slave_equality_constraints}
\bvec{M}_m = \hat{\varepsilon}_0 - \sum_{\alpha \ne m} \bvec{M}_\alpha \;.
\end{equation}
The homogeneous inequality constraints on the slave,
$\bvec{M}_m \circ \bvec{\psi}(i) \ge 0 \; (\forall \,i),$
transform into inhomogeneous inequality constraints that couple the remaining $m-1$ {\em free} particles:
\begin{equation}
\label{Eqn:coupled_inequality_constraints}
\sum_{\alpha \ne m} \bvec{M}_\alpha \circ \bvec{\psi}(i) \le 1 \;.
\end{equation}
We consolidate the $m(m-1)$ degrees of freedom of the free particles
into the supervector $\bvec{M}^{\rm free}$ having
components $(\bvec{M}_1, \bvec{M}_2, \ldots, \bvec{M}_{m-1})$ in $\mathbb{R}^{m(m-1)}$. Optimization then proceeds in $\mathbb{R}^{m(m-1)}$
with the $\bvec{M}^{\rm free}$ restricted by $(m-1)N$ homogeneous inequality constraints from \eqn{discrete_inequality_constraints} with $\alpha < m$ and $N$ inhomogeneous inequality constraints from \eqn{coupled_inequality_constraints}.  The combination of homogeneous and inhomogeneous inequality constraints forms a closed convex polytope that bounds the feasible region.  Each local minimum of $\Phi(M)$ (and thus, the global minimum) lies at a vertex of this polytope~\cite{Korenblum:03}.

An example of the asymmetric $M$-representation for $m=2$ is shown in Fig.\ \ref{fig:cres_mspace_constrained}. In this case there are four bounding constraints: two homogeneous inequality constraints having boundaries passing through the origin and two inhomogeneous inequality constraints (from the slave cluster) with boundaries intersecting at
$\hat{\varepsilon}_0$~\cite{ft:11}.
$\Phi$ is infinite at the polytope vertices at the origin and $\hat{\varepsilon}_0$. The two other vertices correspond to index-permutation-equivalent global minima.

The minimization problem can be visualized and easily solved in this manner only for $m=2$: As $m$ increases the number of polytope vertices, and hence the number of local minima, grows rapidly, and the global minimization problem becomes difficult. Korenblum and Shalloway~\cite{Korenblum:03} solved this by an expensive, random exploration of the vertices.

\subsection{Cluster representatives and the approximate global solution}
\label{section:zeroth_order_solver}
\subsubsection{Representatives}
We take a different approach: Rather than trying to identify the minimizing vertex directly, we exploit the fact that the $m^2$ components of $M$ can be determined by the $m^2$ low-frequency components of an  appropriately chosen subset ${\cal R}= \{r_1, r_2, \ldots,r_m\}$ of $m$ items, which we call \emph{representatives}. To make this explicit we write a matrix analog of \eqn{w_psi_discrete} over $\cal R$ as
\begin{equation}
\label{Eqn:w_mtrue_psi}
 W^{\cal R} = M \circ \Psi^{\cal R} \;,
\end{equation}
where
\begin{subequations}
\begin{eqnarray*}
W^{\cal R}_{\alpha \beta} & \equiv & w_\alpha(r_\beta) \qquad (1 \le \alpha,\beta \le m) \\
\Psi^{\cal R}_{n \alpha} & \equiv & \psi_n(r_\alpha) \quad
\left\{ \! \begin{array}{cccc}
   (1 & \le & \alpha &  \le m) \\
   (0 & \le & n & <  m)
  \end{array} \right. \;, \label{Eqn:psi_R_def}
\end{eqnarray*}
\end{subequations}
and $M$ is the matrix having the $\bvec{M}_\alpha$ as its rows. According to
\eqn{discrete_equality_constraints}, $M$ must satisfy
\begin{equation}
\label{Eqn:M_equality_constraint}
\sum_\alpha M_{\alpha n} = \delta_{n0} \;.
\end{equation}
As shown in Appendix~\ref{appendix:rank_proof}, there always exists at least one subset $\cal R$ such that $\Psi^{\cal R}$ is invertible.  With such a subset we can solve \eqn{w_mtrue_psi} for $M$:
\begin{equation}
\label{Eqn:M_from_W}
 M = W^{\cal R} \bullet (\Psi^{\cal R})^{-1} \;,
\end{equation}
where $\bullet$ denotes the inner product over the cluster index $\alpha$.

The usefulness of \eqn{M_from_W} may be questioned since \emph{a priori} we do not
know any $W^{\cal R}$ exactly. However, any data
set amenable to clustering will have at least one item per cluster that will be strongly assigned in the clustering solution; we call such items \emph{candidate representatives}. If we could select a set of representatives ${\cal R}_c$ containing one candidate representative from each cluster, we could use our approximate foreknowledge of their assignment values at the
solution, $W^{{\cal R}^*_c}$, to approximate $M$ at the solution, $M^*$,
via \eqn{M_from_W}.

For example, if item $i_\alpha$ were a candidate representative for cluster
$\alpha$, its assignment in the clustering solution would
be~\cite{ft:12}
\begin{equation}
\label{Eqn:approx_w}
 w^*_\beta (i_\alpha) \approx \delta_{\alpha \beta} \;.
\end{equation}
By choosing $r_\alpha= i_\alpha$ and making similar choices for the other clusters, we would get
\begin{eqnarray*}
%\label{Eqn:w_approx_w_tilde}
W^{{\cal R}^*_c} \approx I \;.
\end{eqnarray*}
This zeroth-order estimate could be used to approximately solve \eqn{M_from_W} for $M^*$:
\begin{subequations}
\begin{eqnarray}
M^* & = & W^{{\cal R}^*_c} \bullet (\Psi^{{\cal R}_c})^{-1}\\
& \approx &   I \bullet (\Psi^{{\cal R}_c})^{-1} = (\Psi^{{\cal R}_c})^{-1} \equiv M^0\;. \label{Eqn:M0_def}
\end{eqnarray}
\end{subequations}
In agreement with \eqn{M_equality_constraint}, $M^0$ would
satisfy~\cite{ft:13}
\begin{equation}
\label{Eqn:M0_equality_constraint}
\sum_\alpha M^0_{\alpha n} = \delta_{n0} \;.
\end{equation}

Knowing $M^0$ would allow us to define zeroth-order estimates $\bfw^0_\alpha$ for \emph{all} the items via \eqn{w_psi_discrete} with $\bvec{M}_\alpha = \bvec{M}_\alpha^0$, where the $\bvec{M}^0_\alpha$ are the rows of $M^0$:
\begin{equation}
\label{Eqn:w0_psi_discrete}
\bfw^0_\alpha  = \bvec{M}^0_\alpha \circ \bvec{\bfpsi} \;.
\end{equation}
However, the $\bfw^0_\alpha$ would not necessarily satisfy the inequality constraints of \eqn{discrete_w_inequality_constraints}.  If they did, they would solve the optimization problem (see Sec.\ \ref{section:M0_exact}). If they didn't, they would provide a starting point for refining the solution as discussed in Sec.\ \ref{section:constrained_solver}.

\subsubsection{Finding ${\cal R}_c$}
\label{section:finding_Rc}
\eqn{M0_def} implies that we only need to find the representatives to determine $M^0$. This is trivial when $m=2$: The two active inequality constraints [identified by either pair of intersecting bold and bold-dashed lines in Fig.\ \ref{fig:cres_mspace_constrained}] come from the extremal items $r_1$ and $r_2$ of $\bfpsi_1$, i.e., the minimizer and maximizer of $\psi_1(i)$.  Thus, at the solution $w^*_1(r_2)=0$ and $w^*_2(r_1) = 0$, and the equality constraints imply that $w^*_1(r_1)=1$ and $w^*_2(r_2)=1$: $r_1$ and $r_2$ not only generate the active constraints, but are also the representatives, which in this case are perfectly assigned in the solution.
%
%\singlespace
\begin{center}
  \begin{figure*}
    \begin{center}
      %%%\begin{minipage}{0.50\linewidth}
      \begin{math}
        \begin{array}{cccc}
          \subfigure[]{
            \label{fig:spiral_data_set}
            \scalebox{1.0}{
              \includegraphics{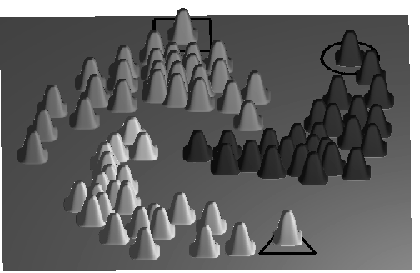}
            }
          }
          & \hspace*{0.0in} &  % Note asymmetry.  This hspace forces
                                % the column to be of some width even
                                % though we don't include it below.
          \subfigure[]{
            \scalebox{1.0}{
              \includegraphics{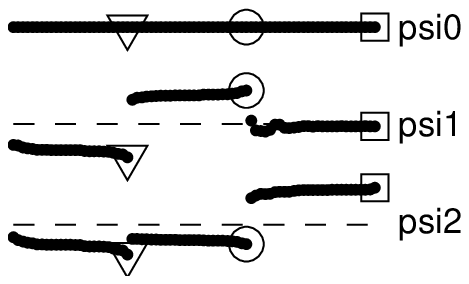}
            }
          } &                   % Eh?  Empty 4th column?  We need to
                                % overlay the \autorightleft arrows
                                % (logically the 2nd column)
                                % on the opaque background of the 4th
                                % column.  If we actually placed it in
                                % the 3rd column, the placement of the
                                % opaque background would occlude it.
                                % So, output the opaque background and
                                % in the 4th column (i.e., after)
                                % place the arrows and move them
                                % far to the left (via hspace)
                                % to appear in the 2nd column.

          \\
          [4ex]\boldmath{\bar{w}^\triangle}\!-\!\textbf{representation} & &
               \boldmath{\bvec{\psi}}^\perp\!-\!\textbf{representation}\\[-2.5ex]
      \subfigure[]{\label{fig:spiral_2d_assignment_polytope}
            \scalebox{1.0}{
              \includegraphics{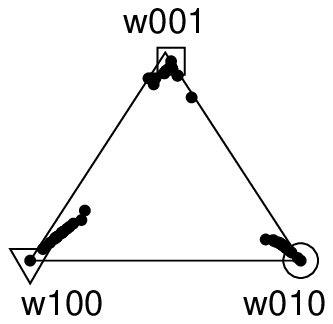}
            }
          }
          &  &
          \subfigure[]{\label{fig:spiral_2d_eig_rep}
            \scalebox{1.0}{
              \includegraphics{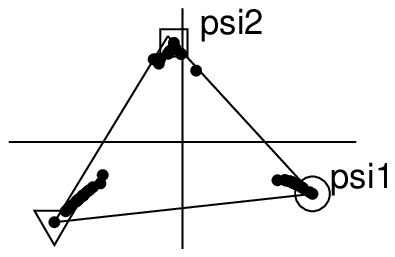}
            }
          }
          &
          \raisebox{0.65in}{
            \hspace{-4.2in}
            \makebox[0pt]{
              %%$\bar{w}^\triangle(i) \autorightleft{$(M^*)^{-1}$}{$M^*$} \bvec{\psi}^\perp(v)$
              $\autorightleft{$(M^*)^{-1}$}{$M^*$}$
            }
          }
          \vspace{-0.10in} \\

          \subfigure[]{\label{fig:spiral_2d_initial_assignment_polytope}
            \scalebox{1.0}{
              \includegraphics{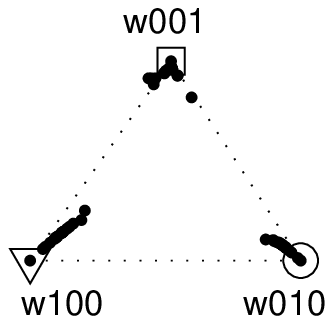}
            }
          }
          & &
          \subfigure[]{\label{fig:spiral_2d_initial_eig_rep}
            \scalebox{1.0}{
              \includegraphics{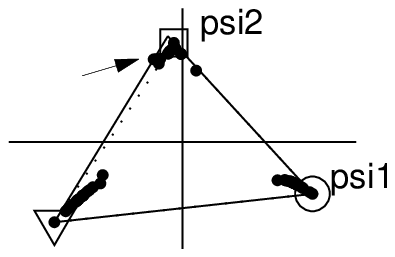}
            }
          } &
          \raisebox{0.65in}{
            \hspace{-4.2in}
            \makebox[0pt]{
              %$\bar{w}^\triangle(i) \autorightleft{$(M^0)^{-1}$}{$M^0$} \bvec{\psi}^\perp(v)$
              $\autorightleft{$(M^0)^{-1}$}{$M^0$}$
            }
          }
          \vspace{-0.10in}
        \end{array}
      \end{math}
\caption{$\bar{w}^\triangle$- and $\protect\bvec{\psi}^\perp$-representations of the spiral problem. The items are represented in the dataspace as peaks with magnitudes determined by their maximal assignment (a),
in the clustering (low-frequency) eigenvector representation (b), in the barycentric coordinates of the $\bar{w}^\triangle$-representation (c) and (e), or in the $\protect\bvec{\psi}^\perp$-representation (d) and (f). Panels (c) and (d) correspond to the refined solution $M^*$, while (e) and (f) correspond to the zeroth-order solution $M^0$. The solid and dotted triangles denote the $M^*$ and $M^0$ feasible region boundaries.  [The solid triangle is superimposed in panel (f) to show how the  triangle expands slightly in the $\protect\bvec{\psi}^\perp$-representation during refinement. The arrow indicates the item that becomes an active constraint in $M^*$.]  The left and right arrows connecting the representations are reminders that $M$ determines the positions of the items in the $\bar{w}^\triangle$-representation and of the triangle vertices in the $\protect\bvec{\psi}^\perp$-representation. [Although it may not be evident in the figure, the points in panels
(c) and (e) and the top vertex in panels (d) and (f) are at slightly different positions.] The different shades of gray in panel (a) denote the hard clustering obtained by quantizing the fuzzy clustering, while the height of a cone shows the strength of the probabilistic assignment of the item to the cluster. The ordering of items in panel (b) was chosen \emph{post facto} to separate the clusters. The dashed lines in this panel are at $\psi_n(i)=0$. The representatives for clusters 1, 2, or 3 are enclosed within triangles, circles, or squares, respectively.
}
    \label{fig:spiral_polytopes}
    \end{center}
  \end{figure*}
\end{center}

The situation is more complicated when $m>2$.  The representatives: (1) may not be maxima and minima of the eigenvectors, (2) may not be the items associated with the active constraints, and (3) may not be perfectly assigned at the solution.  Nonetheless, as discussed above, they will satisfy $w_\alpha(r_\beta) \approx \delta_{\alpha \beta}$ and we will use this property to identify them.

We show how this is done using the $m=3$ spiral problem as an example (\fig{spiral_polytopes}).  Its three low-frequency clustering eigenvectors are shown in panel (b), and the representatives that we would like to find are identified by circles, triangles, and squares. To find ${\cal R}_c$ we imagine that we know $M^*$ and the corresponding assignment vectors $\bfw_\alpha^*$ so that we can map the items into ${\mathbb R}^m$ at the points specified by the 3-vectors $\bar{w}^*(i) \equiv [w^*_1(i), w^*_2(i), w^*_3(i)]$ in panel
(c)~\cite{ft:14}.
Because the $\bar{w}^*(i)$ satisfy the probabilistic equality constraints, these points lie in the 2-dimensional plane that is normal to the vector $(1,1,1)$ and at distance $1/\sqrt{3}$ from the origin. Moreover, they satisfy the probabilistic inequality constraints and thus lie within an equilateral triangle in this plane. (We use ``within'' to include points that lie on the boundary.) This provides barycentric coordinates~\cite{Coxeter:69} in which the three vertices of the triangle correspond to the cluster assignments  $(1,0,0)$, $(0,1,0)$, and $(0,0,1)$; we will call these the $\alpha = 1$, 2, and 3 vertices, respectively. The three components of $\bar{w}^*(i)$ are given by the three distances of point $i$ from the three sides of the triangle.  Thus, if point $i$ lies on the side of the triangle opposing vertex $\alpha$, the inequality constraint $w_\alpha(i) \ge 0$ is active.  We call this the
\emph{$\bar{w}^\triangle$-representation} [panel (c)]. Although it may not be evident in the figure, consistent with the even distribution of active inequality constraints between the clusters (Appendix \ref{appendix:inequality_constraint_proof}), each side of the triangle intersects exactly two items.

The candidate representatives are the items that are close to the three vertices, and we want to choose one from the vicinity of each vertex to compose ${\cal R}_c$. We can do this by choosing the three items that (when taken as vertices) define the triangle of largest area. It is easy to show that the triangular area defined by any subset $\cal R$ of three items located at their solution positions is $| W^{{\cal R}^*}| /(2 \sqrt{3})$.  Thus, we can find a good ${\cal R}_c$ by finding the subset $\cal R$ that maximizes $|W^{{\cal R}^*}|$.

Since we don't actually know $M^*$ or the $\bar{w}^*(i)$, it is not obvious how to proceed.  However, \eqn{w_mtrue_psi} implies that
\begin{eqnarray}
\label{Eqn:Wstar_from_Mstar}
|W^{{\cal R}^*}| = |M^*| \, |\Psi^{\cal R}| \;,
\end{eqnarray}
so, since $M^*$ is fixed (though unknown), selecting the $\cal R$ that maximizes $|W^{{\cal R}^*}|$ is equivalent to selecting the $\cal R$ that maximizes $|\Psi^{\cal R}|$.  This is straightforward because $\Psi^{\cal R}$ does not depend on $M$. Formally, maximizing $|\Psi^{\cal R}|$ is a combinatoric problem that could be solved by comparing the determinants for all subsets $\cal R$. However, this would be exponentially expensive in $N$. Instead we use an efficient greedy algorithm that selects the representatives solely from the subset of candidate representatives.  This may not exactly maximize the determinant, but will be adequate to determine an ${\cal R}_c$ that gives, via \eqn{M0_def}, an $M^0$ that can be used as a starting point for refinement.

We leave the details of the greedy algorithm to Appendix~\ref{appendix:greedy}, but it is useful to establish its geometric framework here, continuing to use the spiral problem as an example: We first plot each item in the $2$-dimensional \emph{$\bvec{\psi}^\perp$-representation} using the $2$-vector $\bvec{\psi}^\perp(i) = [\psi_1(i), \psi_2(i)]$ [panels (d) and (f)]. [No information is lost in this projection from the low-frequency subspace since $\psi_0(i) = 1 \; (\forall \, i)$.]  These vectors are independent of $M$~\cite{ft:15}; rather, in this representation $M$ determines the position of the inequality constraint bounding
triangle.
As explained in Appendix \ref{appendix:bounding_triangle}, the $\bvec{\psi}^\perp$ coordinates of  the three bounding triangle vertices are the columns of the bottom two rows of $M^{-1}$. When $M=M^*$ [panel (d)], the vertices may not coincide with any items, but all the items will lie within the bounding triangle.  When $M=M^0$ [panel (f)], the vertices of the triangle coincide with the representatives, but some items may violate the inequality constraints and lie outside the triangle. (Four items in the upper left corner are outside the triangle in this example.) The greedy algorithm operates within the $\bvec{\psi}^\perp$-representation to identify ${\cal R}_c$.

The approach generalizes easily to higher $m$: The $\bar{w}^*(i)$ are now $m$-vectors. The $\bar{w}^\triangle$-representation is in an $(m-1)$-dimensional hyperplane normal to the vector $(1,1, \ldots,1)$ in ${\mathbb R}^m$ and provides barycentric coordinates for the $\bar{w}^*(i)$. ${\cal R}_c$ is comprised of the subset of $m$ items that, when located at their solution positions in the $\bar{w}^\triangle$-representation, are the vertices of the $(m-1)$-simplex of largest hypervolume. This hypervolume, for any subset $\cal R$, is proportional to $|W^{{\cal R}^*}|$ so, via \eqn{Wstar_from_Mstar}, we can transform the problem of selecting ${\cal R}_c$ to that of finding the
% ${\cal R}$
the $m$ items that maximize $|\Psi^{\cal R}|$. This problem is equivalent to maximizing the hypervolume of the $\bvec{\psi}^\perp$-representation simplex having vertices at $\{\bvec{\psi}^\perp(i):\, i \in {\cal R}\}$.  Once ${\cal R}_c$ has been identified, it is used to determine $M^0$ via \eqn{M0_def}, and $M^0$ is used to determine $\bfw^0_\alpha$ via \eqn{w0_psi_discrete}.

\subsection{Refinement}
\label{section:constrained_solver}
\subsubsection{Case when $M^0$ is the exact solution}
\label{section:M0_exact}
If the $\bfw^0_\alpha$ satisfy all the inequality constraints, they provide the unique solution to the uncertainty minimization problem. To prove this, consider the $\bvec{\psi}^\perp$-representation of an $m=3$ problem where the inequality constraints are satisfied. As in the spiral problem, the representatives are at the vertices of the $\bvec{\psi}^\perp$ triangle determined by $M^0$, and as discussed above, transforming $M^0$ to $M$ moves the sides of this triangle. Moving any side inwards would leave a representative outside the triangle, thus violating an inequality constraint. And, since all points are already within the triangle (i.e., all inequality constraints are satisfied), moving any side outwards would result in that side contacting less than two points, i.e., one of the clusters would have less than the required (Appendix \ref{appendix:inequality_constraint_proof}) $m-1=2$ active inequality constraints. Therefore, in this case $M^*=M^0$ must be the unique solution.  As can be inferred from the analysis of \fig{m_2}, $M^0$ is always the unique solution for $m=2$ problems.

\subsubsection{Linearizing $\Phi(M)$}
If the $\bfw^0_\alpha$ violate any of the inequality constraints, $M^0$ is not a solution but can be used as the starting point for further refinement.  Since it is expected to be near $M^*$, we can expand the objective function in its neighborhood to first-order as

\begin{eqnarray}
\Phi(M) & = & - \sum_\alpha \log \frac{\bvec{M}_\alpha \circ \bvec{M}_\alpha}{\bvec{M}_\alpha \circ \hat{\varepsilon}_0}    \nonumber \\
        & \approx & \Phi(M^0) + \sum_\alpha \left( \bvec{M}_\alpha - \bvec{M}_\alpha^0 \right) \circ \left. \bvec{\nabla}_\alpha \Phi(M) \right|_{M=M^0}\;,
% Put this back for single-column/double-column
% \nonumber\\ &&
\label{Eqn:gradient}
\end{eqnarray}
where
\[
%\label{Eqn:gradient_2}
\bvec{\nabla}_\alpha \Phi(M) \equiv \frac{\delta \Phi(M)}{\delta \bvec{M}_\alpha} =
- 2 \frac{\bvec{M}_\alpha}{|\bvec{M}_\alpha|^2} + \frac{\hat{\varepsilon}_0}{\bvec{M}_\alpha \circ \hat{\varepsilon}_0}
\]
is the gradient of $\Phi(M)$ with respect to $\bvec{M}_\alpha$. Local minimization using this linear approximation and the constraints of \eqns{M_constraints} pose a linear programming (LP) problem, which can be solved by standard methods.

A simple approach would be to: (1) apply LP using \eqn{gradient} and all the constraints to find an improved, constraint-satisfying solution $M^1$, (2) set $M^0 \leftarrow M^1$, and (3) repeat (1) and (2) until sufficient convergence is achieved. This amounts to constrained gradient-descent local minimization. However, we do not expect to encounter the slow convergence problems that sometimes plague gradient descent because all the LP solutions, as well as the true solution, are at vertices of the feasible
polytope~\cite{ft:16}.
Therefore, even the first iteration will drive the solution to a vertex, and the solution will not change at the next iteration unless the vertices are very dense on the scale set by the curvature of $\Phi(M)$. Thus, rapid convergence is expected.

\subsubsection{Reducing the number of constraints included in LP}
However, the cost of standard LP solvers (e.g., simplex and interior point methods) grows rapidly [$O(N_c^{1.5}$)] with the number of constraints $N_c$, which may be
large~\cite{ft:17}.
While there are $mN$ inequality constraints, only $m(m-1)$ of these are active at $M^*$.  These alone need to be included in the LP problem to guarantee that all the inequality constraints will be satisfied. Since we will often be interested in problems where  $m \sim O(10)$ and $N \sim O(10^4)$, it would accelerate the LP solver by multiple orders of magnitude if the number of constraints provided to it were reduced to $O(m^2)$.

We do not know the active constraints \emph{a priori}, but can find them rapidly by an iterative procedure that exploits the fact that (as discussed above) at $M^*$  exactly $m-1$ points will lie on each of the $m$ faces of the bounding simplex in the $\bvec{\psi}^\perp$-representation. To motivate this procedure, consider the refinement of the spiral problem (\fig{spiral_polytopes}). The left side of the (dotted) $M^0$ triangle [panel (f)] must move outwards to include the four points in the upper left region that are excluded from its interior; this motion must leave the side intersecting two points. Because the objective function $\Phi(M)$ constitutes an inward ``pressure'' on the triangle, $M^*$ will correspond to the situation where the smallest expansion that can accomplish this is used. Consequently, the left side will pivot outwards about the lower left corner until it intersects the item identified by the arrow.  Each
side of the resulting $M^*$ triangle [panel (d)] will intersect $m-1=2$ points, and these points will be near (but not identical with) the $m=3$ vertices of the $M^0$ triangle. These six intersections will identify the $m(m-1)=6$ active constraints.

This suggests that, for $m=3$ in general, the two points lying on a side of the $M^*$ triangle will be near different vertices and, subject to this restriction, will be the points that are farthest outside the $M^0$ triangle. This easily generalizes to $m>3$: Each of the $m$ faces of the $M^*$ simplex will contain $m-1$ item points, each near a different vertex. These $m(m-1)$ points are the most likely to lie outside the $M^0$ simplex. Thus, it is sensible to initially attempt a LP solution using only the inequality constraints corresponding to these $m(m-1)$ face-item point pairs. [If point $i$ lies on the face opposing vertex $\alpha$, this face-item pair corresponds to the active inequality constraint $w_\alpha(i) = 0$.]  However, this is only a heuristic argument, and inequality constraints may
still be violated in the partially constrained LP solution. If so, we iterate while adding to an \emph{included constraint list} $\cal C$ (of face-item pairs) the violated constraints that are identified by the above criteria as most likely to be active. The procedure terminates when all the inequality constraints are satisfied.  Termination is guaranteed because inequality constraints are only added to, and never removed from, the included constraint list. The procedure is formalized below.

\subsubsection{Refinement Algorithm}
 \begin{enumerate}
 \item Initialize $\cal C$ to the empty set.
 \item Perform hard clustering based on the $M^0$ assignments: Item $i$ is assigned to the cluster (vertex) $\alpha$ that maximizes $w^0_\alpha(i)$.  We call this subset of items ${\cal S}_\alpha$.
\item
Identify the item (designated $i_\beta$) from ${\cal S}_\beta \; (\beta \ne \alpha)$ that is farthest outside the face opposing vertex $\alpha$.  This identifies the $m-1$ constraints corresponding to the face-item pairs $(\alpha,i_\beta:\,\beta \ne \alpha)$.  As shown in Appendix \ref{appendix:ordering}, the ordering of the item points relative to the simplex faces is the same in the $\bar{w}^\triangle$- and $\bvec{\psi}^\perp$-representations. Therefore, we determine the ordering in the $\bar{w}^\triangle$-representation barycentric coordinates since this is simple: $w_\alpha(i)$ is the distance of an item point $i$ from the $\alpha$-opposing face (positive if inside, negative if outside the simplex). When executed for all $m$ faces this procedure identifies $m(m-1)$ inequality constraints $\cal C'$.
\item $\cal C \leftarrow \cal C \cup \cal C'$.
\item Apply the LP solver with the equality constraints, the inequality constraints in $\cal C$, and the linear objective function approximation of \eqn{gradient}.
\item Check for satisfaction of all inequality constraints and for convergence according to  $\max_{\alpha,i} |\bfw^1_\alpha(i)-\bfw^0_\alpha(i)| < \rho_{\rm LP}$, where $\rho_{\rm LP}$ is a small number, and $\bfw^1_\alpha(i)$ and $\bfw^0_\alpha(i)$ are the values determined by $M^1$ and $M^0$, respectively.  If both conditions are satisfied, terminate with $M^*=M^1$; if not, set $M^0 \leftarrow M^1$ and return to step 2.
 \end{enumerate}

When the algorithm is applied to the spiral problem, $\cal C$ is set to the active constraints in a single
step~\cite{ft:18}.

\section{Overall Computational Algorithm}
\label{section:algorithm}
Combining the steps described in Sec.\ \ref{section:computational}, the overall algorithm is:
\begin{enumerate}
\item Compute and precondition $\Gamma$ as described in Appendix \ref{appendix:precondition}.
\item Compute 20~\cite{ft:19} low-frequency clustering eigenvalues and eigenvectors
using the Lanczos method~\cite{Golub:96a}.
  This is more efficient than computing the full eigensystem, but will converge slowly if the eigenvalues are densely-packed near zero (as they often are). To exclude this possibility we employ a shift-and-invert spectral transformation ~\cite{Lehoucq:00b}, which spreads out the small eigenvalues by transforming them into the large magnitude eigenvalues of a related spectral decomposition having the same eigenvectors.
\item Following Ref.\ \cite{Korenblum:03}, determine $m$ according to the lowest spectral gap satisfying $\gamma_m/\gamma_{m-1} > \rho_\gamma$, where $\rho_\gamma$ is the {\em minimum gap parameter}.
 If there is no gap, the algorithm has determined that there are no clusters and terminates.
\item Identify the representatives and compute the zeroth-order solution $M^0$ and $\bfw^0_\alpha$ using the procedure of Sec.\ \ref{section:zeroth_order_solver}.
\item Determine if $\bfw^0_\alpha$ violates any inequality constraints.  If so, iteratively refine $M^0$ to $M^*$ using the  procedure of Sec.\ \ref{section:constrained_solver} and,  via \eqn{w_psi_discrete}, compute the refined solution  $\bfw^*_\alpha$.  Otherwise, $\bfw^*_\alpha=\bfw^0_\alpha$.
\item Following Ref.\ \cite{Korenblum:03}, test the solution against the minimum certainty conditions
${\overline\Upsilon}_\alpha(M) > \rho_\Upsilon\; (\forall \, \alpha)$, where $\rho_\Upsilon$ is the \emph{minimum certainty parameter}. If it satisfies them, the solution is accepted. If not, the eigenspectrum can be tested for higher spectral gaps, and the algorithm proceeds with step 4.
If desired, the fuzzy solution can be quantized to a hard clustering by assigning item each $i$ to the cluster having the largest assignment value; these hard clusters may be recursively analyzed.
\end{enumerate}

\section{Results}
\label{section:results}
We tested the efficiency of our method for uncertainty minimization by using it for fuzzy spectral clustering of a family of synthetic data sets containing up to $N=20,000$ items. Further, we showed that it can be applied to both symmetric and asymmetric $\Gamma$ matrices popular in the literature.

\subsection{Implementation}
\label{section:implementation}
The C++ implementation was compiled using {\tt gcc} version 4.1.2 and {\tt g77} version 3.3.5 under {\tt -O3} optimization. It accesses low-level LAPACK~\cite{Anderson:99} routines through LAPACK++~\cite{Stimming:06} version 2.5.2, interfaces to the ARPACK~\cite{Lehoucq:98} Lanczos solver through the ARPACK++ C++ wrappers~\cite{Gomes:97}, and solves constrained linear programs using the  GLPK simplex method~\cite{Makhorin:06} version 4.9.  The scaling benchmarks of Sec.~\ref{section:scaling} were executed on a dedicated quad CPU 3.46 GHz Pentium 4, configured with 4 GB of RAM and 4 GB of swap space, and running a 64-bit version of SuSE Linux. The numerical precision parameter was $\epsilon = 2.22045 \times 10^{-16}$. The minimum gap and minimum certainty parameters were set to $\rho_\gamma = 3$ and  $\rho_\Upsilon = 0.68$~\cite{Korenblum:03}. The LP convergence parameter was $\rho_{LP} = 0.001$.

\subsection{Computational efficiency and scaling}
\label{section:scaling}
To evaluate the efficiency and cost scaling of uncertainty minimization, we applied it to synthetic data sets containing from $2$ to $10$ clusters and from $5,000$ to $20,000$ items arranged in a pyramid of blocks in a two-dimensional dataspace.  For these tests we used the Laplacian $\Gamma$ defined by \eqns{dynamical_properties} and the definitions of  $S$ and $D_\pi$ arising from the continuous dynamical interpretation of Ref.\ \cite{Korenblum:03}:
\begin{subequations}
\label{Eqn:Gamma_KS}
\begin{eqnarray}
S_{ij} & = & \frac{e^{-d_{ij}^2/2 \langle d^2_0 \rangle}}
                {d_{ij}^2} \qquad(i \ne j) \\
(D_\pi)_{ii} & = & N^{-1} \label{Eqn:Gamma_KSb} \;,
\end{eqnarray}
\end{subequations}
where $d_{ij}$ is the Euclidean distance between items $i$ and $j$ in the dataspace and $d^2_0$ is a characteristic distance of the problem:
\begin{subequations}
\label{Eqn:d0squared}
\begin{eqnarray}
\langle d^2_0 \rangle & = & N^{-1} \sum_{i=1}^N d_{i <}^2  \label{Eqn:d_0} \\
d_{i<} & \equiv & \min_{j \ne i} d_{ij} \;.
\end{eqnarray}
\end{subequations}

\begin{center}
\begin{figure}
\scalebox{1.0} {
    \includegraphics{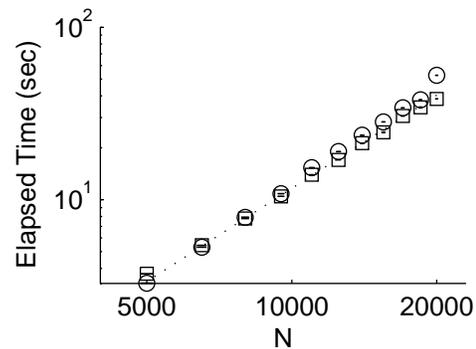}
}
   \caption{Log-log plot of elapsed computational time versus $N$ for synthetic benchmarks.  $N$ was varied from $5,000$ to $20,000$ in steps of $1,500$.  The results shown for $m = 2 \, (\ocircle)$ and $10 \, (\Box)$ are averages over five runs and are representative of those for  $2<m<10$. (The standard errors of the mean are too small to be discernible.) The dotted line is the least-squares linear fit and has slope $1.8$.}
    \label{fig:scaling}
\end{figure}
\end{center}

These problems required up to four invocations of the LP solver, with the number increasing with $m$, but not evidently with $N$. The log-log plot in \fig{scaling} shows that execution time was proportional to $N^{1.8}$ with little dependence on $m$. Execution time was dominated by the calculation of $S$ and by the eigensolver (each having roughly equal cost), with uncertainty minimization contributing  $\lesssim 10 \%$ of the total in all problems tested. The largest problem ($m=10$, $N=20,000$), which is of the scale of biological microarray gene expression data sets, only required about 30 seconds on a commodity processor.

\subsection{General applicability}
\label{section:fcps}
Uncertainty minimization is applicable to spectral clustering using any $\Gamma$ defined by \eqns{dynamical_properties}, including unnormalized and normalized forms that are popular in the literature.  Of course, the success
of any method will depend on the choices of $S$ and $D_\pi$, which are highly problem-specific, and we do not address this issue here. Our goal was to demonstrate the applicability of uncertainty minimization to this wide range of formulations.
Thus, in addition to the tests described above using the $\Gamma$ of \eqns{Gamma_KS}, we applied uncertainty minimization to the spiral problem using two other forms of $\Gamma$. The first one, an asymmetrically normalized Laplacian $\Gamma$  (\cite{Chung:97} for review) with  $S_{ij}$ a Gaussian function of $d_{ij}$, commonly arises when a Markovian~\cite{Belkin:03, Weber:04, Nadler:05, Nadler:06} rather than a continuous~\cite{Korenblum:03} dynamical interpretation is used. It is specified by \eqns{dynamical_properties} with
\begin{subequations}
\label{Eqn:Gamma_asym_normalized_laplacian}
\begin{eqnarray}
S_{ij} & = & e^{-d_{ij}^2/2 \sigma^2} \label{Eqn:S_asym}\\
(D_{\pi})_{ii} & = & \frac{\sum_j S_{ji}}{\sum_{jk} S_{jk}} \label{Eqn:Gamma_asym_normalized_laplacianb} \;,
\end{eqnarray}
\end{subequations}
where $\sigma$ is chosen by empirical tuning~\cite{Belkin:03, Weber:04, Nadler:05, Nadler:06} or  heuristics~\cite{ZelnikManor:04,vonLuxburg:07}. We chose $\sigma^2=\langle d_0^2 \rangle$. (This type of $\Gamma$, but with a non-Gaussian $S$, also frequently arises in image segmentation~\cite{Shi:97, Meila:00, Meila:01, Malik:01} where it is motivated by the ``normalized cut'' variant of the min-cut graph partitioning method~\cite{Shi:97}.)
We also tested the symmetric, unnormalized Laplacian form (\cite{Mohar:91} for review) of $\Gamma$ specified by
\begin{subequations}
\label{Eqn:Gamma_unnormalized_laplacian}
\begin{eqnarray}
S_{ij} & = & e^{-d_{ij}^2/2 \sigma^2} \label{Eqn:S_familiar}\\
(D_\pi)_{ii} & = & N^{-1}\;.
\end{eqnarray}
\end{subequations}
 This form is popular in graph partitioning problems (e.g., VLSI circuit partitioning~\cite{Hagen:92, Chan:93b, Alpert:99}, parallel matrix factorization~\cite{Pothen:90}, and computational load balancing~\cite{Barnard:94, Hendrickson:95}), where it is used to approximate the solution to the ``ratio cut'' variant of the min-cut graph partitioning method~\cite{Hagen:92}.  When applied to graph partitioning the $S_{ij}$ are simply edge weights, but to apply it to the spiral data clustering problem the $S_{ij}$ must be computed from the $d_{ij}$; for this we again used the Gaussian form of \eqn{S_familiar} because it is popular in dataspace clustering~\cite{vonLuxburg:07}.

\begin{center}
\begin{figure*}
%%%[htbp]
  \begin{center}
      \subfigure[]{\label{fig:MDC_spiral}
        \scalebox{1.0}{
          \includegraphics{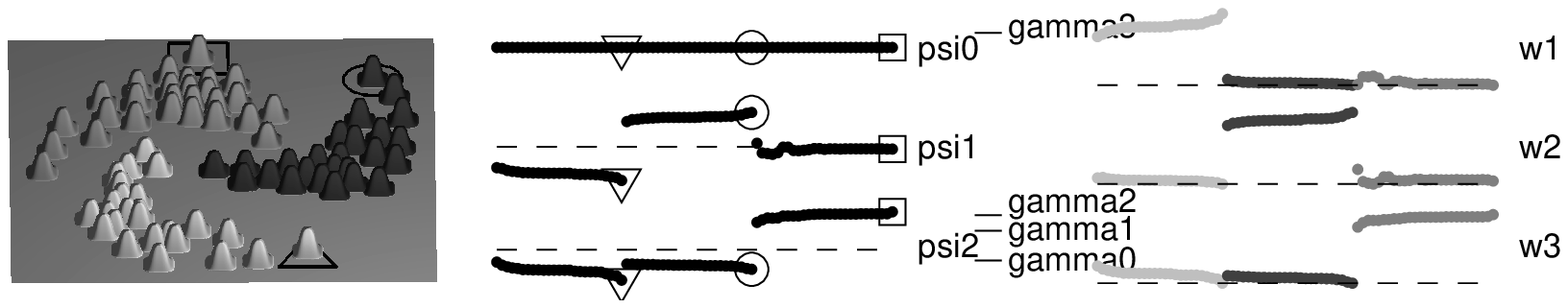}
        }
       }
      \subfigure[]{\label{fig:fuzzy_asymmetric_normalized_laplacian_spiral} \scalebox{1.0}{
          \includegraphics{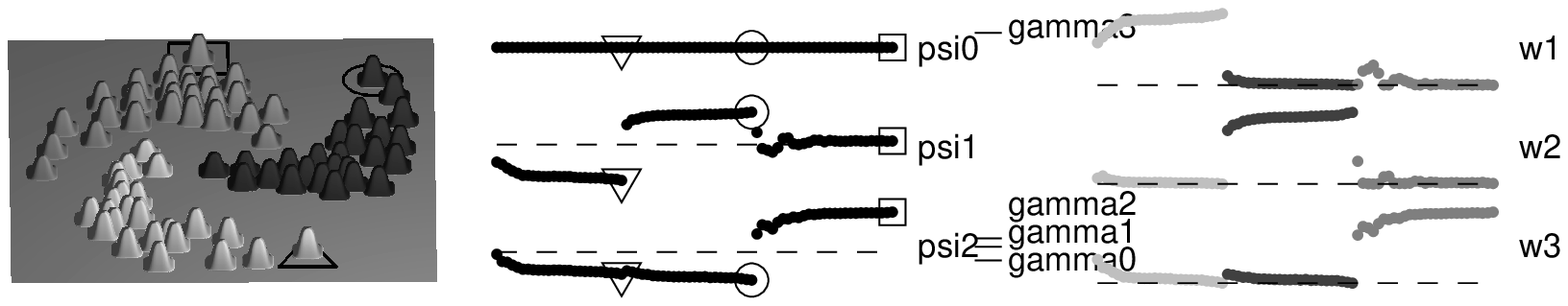}
                                       }}
      \subfigure[]{\label{fig:fuzzy_unnormalized_laplacian_spiral} \scalebox{1.0}{
          \includegraphics{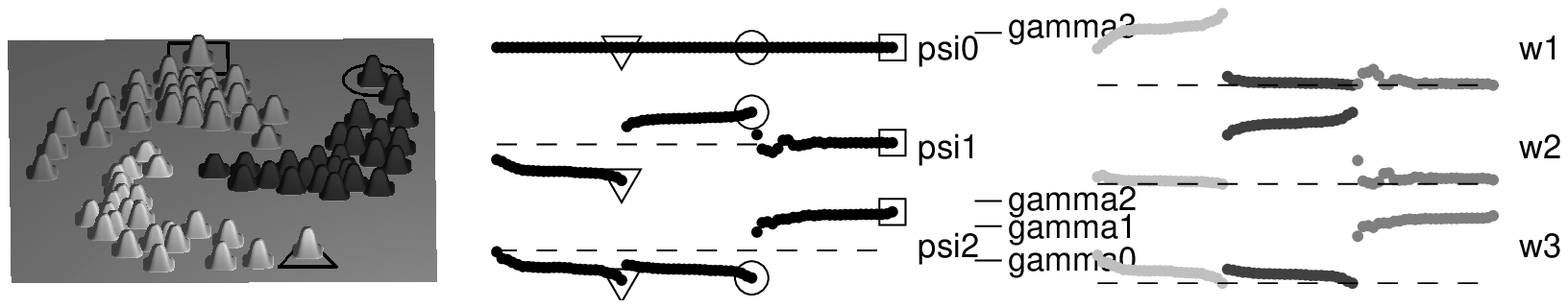}
                                       }}
      \subfigure[]{\label{fig:fuzme_spiral} \scalebox{1.0}{
          \includegraphics{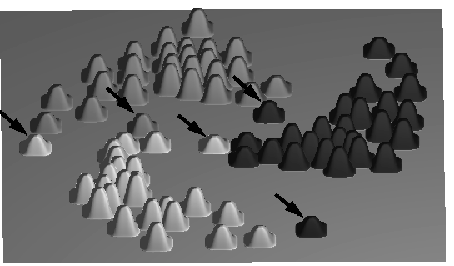}
                                       }}
\caption{Fuzzy spectral clustering by uncertainty minimization of the spiral problem using three different $\Gamma$ matrices. The fuzzy clusterings and clustering eigenvectors, eigenvalues, and assignment vectors computed using the $\Gamma$ matrices specified in (a) \eqns{Gamma_KS}, (b) \eqns{Gamma_asym_normalized_laplacian}, and (c) \eqns{Gamma_unnormalized_laplacian} are shown. Representatives are indicated by triangles, circles, and squares in the two left columns. (d) Application of fuzzy $k$-means with extragrades~\cite{Degruijter:88} to this problem; arrows identify misclassified items.}
    \label{fig:spiral_plots}
  \end{center}
\end{figure*}
\end{center}

\fig{spiral_plots} shows the results obtained by using uncertainty minimization for fuzzy spectral clustering of the spiral problem with the $\Gamma$ matrices defined by \eqns{Gamma_KS}, \eqnref{Gamma_asym_normalized_laplacian}, and \eqnref{Gamma_unnormalized_laplacian}.
In each case the algorithm selected the same three representatives and the LP solver was invoked twice. While there were minor differences in the $\bfw_\alpha$ along the cluster boundaries, the use of all three $\Gamma$ gave essentially the same results.  In contrast [panel (d)], the spiral problem confounded $k$-means with ``extragrades'' (which is an outlier-robust variant of $k$-means)~\cite{Degruijter:88}. As discussed in the Introduction, this failure of $k$-means is not surprising, given the irregular, interlocking nature of the clusters.

\section{Discussion}
\label{section:discussion}

To date, spectral clustering has been used primarily for hard partitioning. Prior studies~\cite{Korenblum:03,Weber:04} have suggested that fuzzy spectral clustering could be accomplished by using the $m$ low-frequency eigenvectors of $\Gamma$ as a linear basis for expanding, via a transformation matrix $M$, the fuzzy cluster assignment vectors $\bfw_\alpha$, where
$w_\alpha(i)$ is the probability that item $i$ is assigned to cluster $\alpha$. Korenblum and Shalloway~\cite{Korenblum:03} suggested that $M^*$, the optimal $M$, is best identified by uncertainty minimization, which minimizes the probabilistic overlap between clusters. Uncertainty minimization has the additional advantage of providing measures (the final values of the objective function and fractional cluster certainties) that quantify the quality of a clustering, which can be as important as the clusterings themselves.  However, Korenblum and Shalloway did not provide an efficient means of solving this challenging non-convex global minimization problem, which limited their approach to small data sets with $N \sim O(10^2)$ items.  Alternatively, Weber et al.\ \cite{Weber:04} suggested that $M$ could be determined by perturbative approximation from almost-block-diagonal matrices, but this approach gives $\bfw_\alpha$ that only approximately satisfy the probabilistic constraints of \eqns{w_constraints}. Thus, until now there has been no computationally practical, exact fuzzy spectral data clustering method.

To address this need we developed an efficient method for uncertainty minimization, which extends the number of items that can be clustered by at least two orders of magnitude: data sets with $N\sim O(10^4)$ can now be analyzed within $\sim 30$ seconds on a commodity processor. Using tests with synthetic data sets having up to $20,000$ items and ten clusters we showed that computational cost scaled $\sim O(N^{1.8})$ and was insensitive to the number of clusters. This implies that as many as $N \sim O(10^6)$ items can be clustered in modest time on a serial machine. The additional cost of uncertainty minimization was small compared to costs common to all spectral clustering methods (e.g., computing $\Gamma$ from the $d_{ij}$ and computing its low-frequency eigensystem).

In developing this approach we elucidated the underlying structure of the uncertainty minimization problem. This revealed  fundamental relationships between four different geometric representations: the $m$-dimensional symmetric $M$-representation, the $m(m-1)$-dimensional asymmetric $M$-representation, and the $(m-1)$-dimensional $\bar{w}^\triangle$- and $\bvec{\psi}^\perp$-representations. All are formally equivalent, but each has advantages: The symmetric $M$-representation has the most direct connection to the minimization problem.  The asymmetric $M$-representation provides a closed feasible region; it is used to prove that all local minima are at polytope vertices and that the inequality constraints are evenly distributed between the clusters at these points.  The $\bar{w}^\triangle$-representation provides barycentric coordinates and makes it evident that the $m$ cluster representatives in ${\cal R}_c$ are those items that determine the $(m-1)$-simplex of largest hypervolume. The $\bvec{\psi}^\perp$-representation motivates the greedy algorithm used to approximate ${\cal R}_c$, which in turn yields $M^0$, the starting point for refinement to $M^*$.

\begin{center}
\begin{figure}
%%%[htbp]
  \begin{center}
    \mbox{
      \subfigure[]{\label{fig:erf_overlap} \scalebox{1.0}{
          \includegraphics{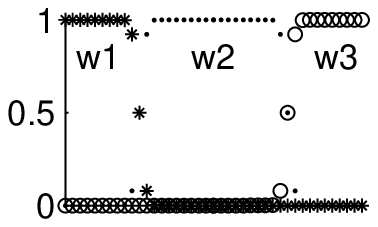}
                                       }}
      \subfigure[]{\label{fig:cos_overlap} \scalebox{1.0}{
          \includegraphics{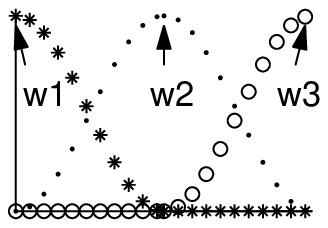}
                                       }}
      }
    \caption{Assignment vectors for a three-cluster problem for eigenvectors with or without a ``level structure.'' Eigenvectors arising from almost-block-diagonal $\Gamma$ have a level structure leading to ``almost-hard'' assignment vectors such as those shown in panel (a). (Different symbols are used for the three assignment vectors.) When there is no level structure, the assignment vectors are much softer, as in panel (b).  However, even in this case representatives (identified by arrows) exist.}
    \label{fig:overlap_plots}
  \end{center}
\end{figure}
\end{center}

The greedy algorithm we used is almost identical to the ``inner simplex method'' used in the Perron Cluster Cluster Analysis method~\cite{Weber:03,Weber:04} for approximate fuzzy data clustering~\cite{ft:20}.
However, our motivation for the algorithm, and consequently our understanding of its domain of validity, are different. The inner simplex method was motivated by earlier studies~\cite{Deuflhard:00,Schuette:03} on perturbation theory of block-diagonal matrices~\cite{Stewart:84}.  These studies exploited two observations: (1) that the $\Gamma$ of well-separated clusters can be brought into almost-block-diagonal form, and (2) that the low-frequency eigenvectors of such a $\Gamma$ are perturbed only in second-order in the non-block-diagonal terms, and therefore, to this order, possess a ``level structure'' in which their components are concentrated near $m$ different values. The inner simplex method aims at finding one item from each level set and thus, in principle, depends on their existence. In contrast, the analysis presented here makes no assumptions about level structure and only presumes that at least one item (i.e., the
representative) can be well-assigned to each cluster.  An example where representatives exist, even though matrix perturbation theory is no longer applicable and the eigenvectors do not have a level structure, is illustrated in \fig{overlap_plots}. Even in this case it is evident that there are three fuzzy clusters, although many of the items will have weak assignments. Thus, the greedy algorithm is more generally applicable than previously stated.

In the two-cluster case the $M^0$ solution is always exact, but in the tested $m>2$ problems, it always violated some of the inequality constraints required for a probabilistic interpretation of the $\bfw_\alpha$. These violations were removed by refinement. The corrections changed $w_\alpha(i)$ by $<0.05$; so, except when high accuracy is needed, the most important role of the refinement may be to provide a rational method for ensuring that the $\bfw^0_\alpha$ satisfy the probabilistic constraints.

Deuflhard and Weber~\cite{Deuflhard:05} used a metastability objective function for clustering protein conformations collected from molecular dynamics simulations that is closely related to the sum of the fractional cluster certainties ${\overline\Upsilon}_\alpha(M)$ defined in \eqn{upsilon}. Their objective function is the sum of terms
\begin{eqnarray*}
\bar{\upsilon}_\alpha(M;t) \equiv \frac{\langle \bfw_\alpha  | e^{-\Gamma \, t} | \bfw_\alpha \rangle}{ \langle \bfl | \bfw_\alpha \rangle} \;,
\end{eqnarray*}
where $t$ denotes a time period which, in practice, is set to a multiple of the molecular dynamics integration time step~\cite{Kube:07, Noe:07}. $\bar{\upsilon}_\alpha(M;t)$ measures the fractional persistence of probability within subregion $\alpha$ of conformation space after stochastic evolution for time $t$, and is identical to the  ${\overline\Upsilon}_\alpha(M)$ except for the presence of the Markov matrix $e^{-\Gamma \, t}$, generated by a $\Gamma$ derived from the molecular dynamics data. Thus, ${\overline\Upsilon}_\alpha(M)$ is the $t \to 0$ limit of $\bar{\upsilon}_\alpha(M;t)$.  It is not clear if a $t$-dependent objective function is appropriate for clustering data that does not arise in a dynamic manner, though this may be worth considering.

Another potentially interesting objective function is the determinant of $M$.  It is intriguing because of its simple geometric interpretation:  We can show that maximizing $|M|$ is equivalent to maximizing the hypervolume of the $(m-1)$-simplex formed in the $\bar{w}^\triangle$-representation by \emph{any} subset of $m$
items~\cite{ft:21}.
This property is attractive since we expect a good clustering to spread the items out in this barycentric representation as much as possible. However, $|M|$ does not have a simple information-theoretic interpretation as does $\Phi(M)$, defined in \eqn{minimum_uncertainty}: $\exp [-\Phi(M)]$ is the product of the fractional cluster certainties, $\overline{\Upsilon}_\alpha$, which are normalized to unity when the corresponding cluster is completely hard, but   $|M|$ does not provide a measure of cluster hardness. Moreover, while optimization using either $|M|$ or $\Phi(M)$ tends to minimize overlap, optimization of $|M|$ also tends to equalize the size of the clusters~\cite{ft:21}. Although this is not necessarily desirable for data clustering, it may be of value in graph partitioning applications that seek to balance partition sizes~\cite{Shi:97,Hagen:92}.

Uncertainty minimization and the method for efficiently solving it presented here are applicable to the wide range of popular $\Gamma$ matrices that satisfy \eqns{dynamical_properties}.  To demonstrate this, we applied uncertainty minimization to the spiral data set, a convenient two-dimensional example with visually-discernible irregularly-shaped clusters, using one asymmetric and two symmetric forms of $\Gamma$. The resulting fuzzy spectral clustering gave similar results with all three $\Gamma$ matrices, while $k$-means did not provide a valid clustering. Of course, these particular forms may not be suitable for all data sets---as in hard spectral clustering, $\Gamma$ must often be tailored to the problem.  Our goal here was to demonstrate the ability of uncertainty minimization to efficiently fuzzify spectral clustering methods. It can now be applied to a wide variety of problem-specific domains, such as those noted in the Introduction.

\section*{Acknowledgments}
The authors are grateful to Sally McKee for the use
of computational resources and to Vince Weaver for help
in administering them.  Partial support was provided for B.S.W.
by DOE and administered by The Krell Institute, Ames, IA.

\appendix
\section{$\Gamma$ Preconditioning}
\label{appendix:precondition}
Numerical errors in computing the eigensystem increase with $\gamma_{N-1}/\gamma_1$ and,  if this ratio is too large, can obscure differences between very small eigenvalues and obfuscate the spectral gap. This can occur if two items within a cluster are exceedingly close (and hence communicate very rapidly) or if clusters are nearly isolated (and hence communicate very slowly). The latter situation can also occur if the data contain outliers---items that are distant from the bulk of the items.  We avoid these problems by preconditioning $\Gamma$ and, at the same time, improve computational efficiency by sparsifying it (i.e., by setting very small transition rates exactly to zero). [This reduces memory requirements and improves cache performance and eigensolver efficiency so that MDC may be practically applied to large problems.  For example, for the largest of the scaling benchmarks considered in Section V.B. (i.e., $m = 10,\; N = 20,000$), the sparsified $\Gamma$ matrix held less than 650,000 independent elements, representing a storage reduction of a factor of $\sim 300$.] This involves three steps: (1) determine appropriate upper ($d_{\rm hi}$) and lower ($d_{\rm lo}$) bounds on the $d_{ij}$, (2) sparsify $\Gamma$ using $d_{\rm hi}$ and check for any resultant graph disconnections, and (3) evaluate the remaining matrix elements and truncate the large-magnitude elements using $d_{\rm lo}$, compute $\gamma_1$, bound $\gamma_{N-1}$, and confirm that $\gamma_{N-1}/\gamma_1$ is properly constrained. If it is not, $d_{\rm lo}$ is increased so that it will be. (Increasing $d_{\rm lo}$ was not required for the examples in this paper, but this step is included as a precaution.)

To avoid excessive numerical error, we want to adjust $\Gamma$ so that
\begin{equation}
\label{Eqn:no_eigenvalue_degeneracies}
\frac{\Delta_\gamma}{\gamma_1} \le \alpha \;,
\end{equation}
where $\Delta_\gamma$ is the expected computational error in the eigenvalues and $\alpha$ is the desired fractional precision, e.g., $\sim O(10^{-2})$.  Typically~\cite{Lehoucq:98, Anderson:99}
\begin{eqnarray*}
\Delta_\gamma \le \epsilon \, \gamma_{N-1}\;,
\end{eqnarray*}
where $\epsilon$ is machine precision.  So \eqn{no_eigenvalue_degeneracies} will be satisfied if
\begin{equation}
\label{Eqn:eigenvalue_condition}
\frac{\gamma_{N-1}}{\gamma_1}\le \alpha/\epsilon \;.
\end{equation}
We expect that $\gamma_{N-1}/\gamma_1$ will depend on $|\Gamma_{\rm hi}|/|\Gamma_{\rm lo}|$, the ratio of the largest to the smallest non-zero $|\Gamma_{i \ne j}|$, and one way to satisfy \eqn{eigenvalue_condition} would be to limit this ratio.  However, when clustering data, e.g., as in the examples of this paper, computing the $\Gamma_{i\ne j}$ from the  $d_{ij}$ constitutes a significant fraction of total cost because exponentiation is required [at least for forms of $S$ in \eqns{Gamma_KS}, \eqnref{Gamma_asym_normalized_laplacian}, and \eqnref{Gamma_unnormalized_laplacian}] and this is wasted for the large fraction of the $\Gamma_{i \ne j}$ that are zeroed during preconditioning.  Therefore, instead of directly limiting $|\Gamma_{\rm hi}|/|\Gamma_{\rm lo}|$, we gain the same result by limiting $d_{\rm hi}/d_{\rm lo}$, the ratio of the largest to the smallest $d_{ij}$.  This allows us to sparsify before evaluating all but a few matrix elements. This indirect approach is not needed when applying uncertainty minimization to spectral clustering of graphs where $S$ is specified \emph{a priori} and, hence, all elements of $\Gamma$ can be inexpensively computed.

\subsection{Determining $d_{\rm hi}$ and $d_{\rm lo}$}
Although a rigorous \emph{a priori} bound on $\gamma_{N-1}/\gamma_1$ depends on $N$ as well as on $|\Gamma_{\rm hi}|/|\Gamma_{\rm lo}|$, we expect that in most cases the two ratios will be roughly of the same order-of-magnitude since $|\Gamma_{\rm hi}|$ and $|\Gamma_{\rm lo}|$ set the scales of the fastest and slowest dynamical processes in the system~\cite{ft:22}.
Thus, we can hope to satisfy \eqn{eigenvalue_condition} by requiring that
\begin{equation}
\label{Eqn:Gamma_bound_condition}
\frac{|\Gamma_{\rm hi}|}{|\Gamma_{\rm lo}|} = \alpha/\epsilon  \qquad \mbox{(not used)}\;.
\end{equation}
However, when $\Gamma$ is asymmetric [i.e., $(D_\pi)_{ii}  \ne 1/N$ as in \eqn{Gamma_asym_normalized_laplacian}], then even this requirement can not be imposed until $D_\pi$ is evaluated, and this would require costly evaluation of all the $\Gamma_{i \ne j}$ prior to sparsification. Thus, instead we apply  \eqn{Gamma_bound_condition} to $\Gamma^S$:
\begin{equation}
\label{Eqn:GammaS_bound_condition}
\frac{|\Gamma^S_{\rm hi}|}{|\Gamma^S_{\rm lo}|} = \alpha/\epsilon\;.
\end{equation}
We expect this to be adequate because in most cases multiplying by $D_\pi^{-1}$ will result in $|\Gamma_{\rm hi}|/|\Gamma_{\rm lo}| < |\Gamma^S_{\rm hi}|/|\Gamma^S_{\rm lo}|$. (This indeed is the case for the examples we have considered.) However, exceptional sets of $d_{ij}$ can be constructed where it will not, so this is not guaranteed.  Nonetheless, we use \eqn{GammaS_bound_condition} because of its reduced cost and the guarantee that \eqn{eigenvalue_condition} will ultimately be satisfied by the confirmation and possible iteration steps described in Sec.\ \ref{section:truncating}.

To minimize the effect of preconditioning on the rest of the eigensystem, we multiplicatively center $|\Gamma^S_{\rm hi}|$ and $|\Gamma^S_{\rm lo}|$ around $|\Gamma^S_{\rm mid}|$, a typical midrange rate. That is, we require
\begin{equation}
\label{Eqn:midway}
\frac{|\Gamma^S_{\rm mid}|}{|\Gamma^S_{\rm lo}|} = \frac{|\Gamma^S_{\rm hi}|}{|\Gamma^S_{\rm mid}|} \;.
\end{equation}
We determine $|\Gamma^S_{\rm mid}|$ by noting that $|\Gamma^S_{i>}|$, the magnitude of the largest $\Gamma^S_{i \ne j}$ in row $i$, is the largest transition rate connecting $i$ to other items. Thus, the median of the $|\Gamma^S_{i>}|$ is a reasonable choice for $|\Gamma^S_{\rm mid}|$.  Because $|\Gamma^S_{i\ne j}|$ depends monotonically on $d_{ij}$ [e.g., see \eqns{Gamma_KS}, \eqnref{Gamma_asym_normalized_laplacian}, and \eqnref{Gamma_unnormalized_laplacian}], this is equivalent to $|\Gamma^S_{\rm mid}| = |\Gamma^S({\rm med}\{ d_{i<}\})|$, where ${\rm med}\{d_{i<}\}$ is the median of the $\{d_{i<}\}$, the smallest off-diagonal elements in each row of the $d_{ij}$ matrix. Thus, determining $|\Gamma^S_{\rm mid}|$ requires computing only one element of $\Gamma^S$.  Once this has been done, \eqns{GammaS_bound_condition} and \eqnref{midway} can be combined to give
\begin{subequations}
\label{Eqn:transition_conditions}
\begin{eqnarray}
\Gamma^S_{\rm lo} & = & |\Gamma^S_{\rm mid}|  \sqrt{\epsilon/\alpha} \label{Eqn:Gamma_lo_condition} \\
\Gamma^S_{\rm hi} & = & |\Gamma^S_{\rm mid}| \sqrt{\alpha/\epsilon} \label{Eqn:Gamma_hi_condition} \;.
\end{eqnarray}
\end{subequations}
We then numerically invert $\Gamma^S_{ij}(d_{ij})$ [e.g., using one of \eqns{Gamma_KS}, \eqnref{Gamma_asym_normalized_laplacian}, or \eqnref{Gamma_unnormalized_laplacian}] with $\Gamma^S_{ij} \to -|\Gamma^S_{\rm lo}|$ and $\Gamma^S_{ij} \to -|\Gamma^S_{\rm hi}|$ to determine $d_{\rm hi}$ and $d_{\rm lo}$, respectively.

\subsection{Sparsification and connected component analysis}
$\Gamma^S$ is sparsified by setting all off-diagonal elements having magnitudes less than $\Gamma^S_{\rm lo}$ to zero. That is,
\begin{equation*}
\Gamma^S_{i\ne j} \to 0 \qquad (\mbox{if } d_{ij}> d_{\rm hi}) \;.
\end{equation*}
To test if this disconnects the graph, we perform a standard connected component analysis~\cite{Cormen:01a}. This initially assigns items to individual sets and then iteratively merges sets whenever any of their respective members are connected.  If distinct subsets (i.e., disconnected components) remain at the end, the algorithm creates hard assignment vectors identifying them. (This process may remove outliers.) Larger subsets may be analyzed as new clustering problems of their own.

\subsection{Truncation and checking the eigenvalue range}
\label{section:truncating}
Having sparsified the (typically large) fraction of insignificantly small off-diagonal elements, we now evaluate the remaining $\Gamma^S_{i \ne j}$ while truncating their maximum magnitudes using
\begin{equation*}
\Gamma^S_{i \ne j} \to -|\Gamma^S_{\rm hi}| \qquad (\mbox{if } d_{ij} < d_{\rm lo}) \;,
\end{equation*}
and compute $D_\pi$ and $\Gamma$.  We can then compute $\gamma_1$ using a Lanczos solver (see Sec.\ \ref{section:algorithm}) and bound $\gamma_{N-1}$ using the Gershgorin Circle Theorem~\cite{Golub:96} and \eqns{diagonal_S} and \eqnref{probability_conservation} to
\begin{equation}
\label{gamma_hi_bound}
\gamma_{N-1} \le 2 \max |\Gamma_{ii}| \;,
\end{equation}
If $\gamma_1$ and the Gershgorin bound on $\gamma_{N-1}$ satisfy \eqn{eigenvalue_condition}, then preconditioning is complete. If not, $d_{\rm lo}$, and hence $|\Gamma^S_{\rm hi}|$, is adjusted so that it will be satisfied when the $|\Gamma^S_{i \ne j}|$ are truncated to the new bound and $\Gamma$ is recomputed. Preconditioning is now complete.

\section{Various proofs}
\subsection{Even distribution of active inequality constraints}
\label{appendix:inequality_constraint_proof}
We prove here that each cluster must be constrained by exactly $m-1$ inequality constraints at each local minimum of $\Phi$ in the feasible region. Consider a local minimum $\bvec{M}^{\times{\rm free }}$ in the asymmetric $M$-representation discussed in Sec.\ \ref{section:geometric}.  Korenblum and Shalloway~\cite{Korenblum:03} have already proved that this must be at a vertex of the feasible polytope. The coordinates at the local minimum of the individual free particles, $\bvec{M}_\alpha^\times \; (1 \le \alpha < m)$, satisfy the inequality constraints of \eqn{discrete_inequality_constraints}, but their homogeneity means that they will also be satisfied for any $\xi_\alpha \bvec{M}_\alpha$ with $\xi_\alpha > 0$. Thus, the free particle inequality constraints acting alone leave the $m-1$ degrees of freedom $\xi_\alpha$ unspecified and are inadequate to force $\bvec{M}^{\times{\rm free }}$ to be at a vertex of the feasible polytope.  Therefore, at least $m-1$ additional active constraints must come from the inhomogeneous inequality constraints associated with the slave particle [\eqn{coupled_inequality_constraints}]. However, the choice of the slave particle in \eqn{slave_equality_constraints} is arbitrary.  Therefore, {\em every} particle must have at least $m-1$ active inequality constraints.  But since only $m(m-1)$ inequality constraints are active at a vertex, each of the $m$ particles must have \emph{exactly} $m-1$ inequality constraints active.  This proof extends to every vertex of the feasible polytope except for those vertices where at least one of the $\bvec{M}_\alpha^\times =0$ (since multiplying such an $\bvec{M}_\alpha^\times$ by $\xi_\alpha$ has no effect). This proof does not preclude the possibility that a single item may be associated with multiple active constraints; i.e., it is possible that $w_\alpha(i)=0$ and $w_\beta(i)=0$ are both active constraints. \{This is the case for the solution to the spiral problem [\figs{spiral_polytopes}(c) and (d)] where $w_2(r_1)=0=w_3(r_1)$ and also $w_1(r_2)=0=w_3(r_2)$.\}

\subsection{Invertibility of $\Psi^{\cal R}$}
\label{appendix:rank_proof}
We prove here that there is at least one subset of $m$ items ${\cal R}$ such that $\Psi^{\cal R}$ is invertible.  We define the $m \times N$ matrix $\Psi$ by $\Psi_{ni} \equiv \psi_n(i) \; (0 \le n < m; 1 \le i \le N)$.  Since its $m$ rows (i.e., the low-frequency eigenvectors)
are linearly independent, $\Psi$ has rank $m$. Therefore, $\Psi$ also has at least $m$ linearly-independent columns.  If the items corresponding to these columns are selected to comprise ${\cal R}$,
then the $m \times m$ matrix $\Psi^{\cal R}$ has full
rank and is therefore invertible.

\subsection{Invertibility of $M$}
\label{appendix:invertibility_M}
We prove here that each $M$ corresponding to a local minimum of $\Phi$ within the feasible region is invertible. As proved in Appendix \ref{appendix:inequality_constraint_proof}, at any such minimum each cluster has $m-1$ active inequality constraints: $m-1$ items lie on each of the $m$ faces of the bounding simplex in the $\bar{w}^\triangle$-representation. Consider a subset $\cal R$ that contains one item from each face. It defines an $(m-1)$-simplex (inscribed within or identical to the bounding simplex) with non-zero hypervolume. This hypervolume is proportional to $|W^{\cal R}|$, implying that $|W^{\cal R}| \ne 0$ and, with \eqn{M_from_W}, implying that $|M| \ne 0$.  Thus, $M$ is invertible.

Actually, the proof holds for every $M$ having all $\bvec{M}_\alpha \ne 0$ that lies at a vertex of the feasible polytope in the asymmetric $M$-representation since  Appendix \ref{appendix:inequality_constraint_proof} applies to all such $M$, not only those at local minima.

\subsection{The bounding simplex in the $\protect\bvec{\psi}^\perp$-representation}
\label{appendix:bounding_triangle}
Analogously to \eqn{w_mtrue_psi}, we may write
\begin{equation}
\label{Eqn:vertex_equation}
W^{\rm vert} = M \circ \Psi^{\rm vert}\;,
\end{equation}
where the columns of $\Psi^{\rm vert}$ are the coordinates of the bounding simplex vertices in the low-frequency eigenvector representation and $W^{\rm vert}$ is the matrix whose rows are the coordinates of the vertices in the $\bar{w}^\triangle$-representation; i.e., $W^{\rm vert}=I$. Inverting this gives $\Psi^{\rm vert} =M^{-1}$. When $M=M^0$, \eqns{M0_def} and \eqnref{vertex_equation} imply that $\Psi^{\rm vert}=\Psi^{{\cal R}_c}$, which is consistent with the zeroth-order placement of the representatives at the vertices.  When $M=M^*$, the vertices may not correspond to item locations, but, as proved in Appendix \ref{appendix:invertibility_M}, $M^*$ is invertible, so $\Psi^{\rm vert} =(M^*)^{-1}$. In both cases, the simplex vertex coordinates in the $\bvec{\psi}^\perp$-representation are given by the columns of $\Psi^{\rm vert}$ with the first row omitted.  (Just as for $\Psi^{\cal R}$, all elements in the first row of $\Psi^{\rm vert}$ are one for any invertible $M$, in particular, for $M^0$ and $M^*$~\cite{ft:23}.)

\subsection{Same ordering of item points in the $\bar{w}^\triangle$- and $\protect\bvec{\psi}^\perp$-representations}
\label{appendix:ordering}
To simplify the proof of identical ordering, we use the spiral problem illustrated in \fig{spiral_polytopes} as a specific example; the proof is easily generalized. We index the vertices in the $\bar{w}^\triangle$-representation as described in Sec.\ \ref{section:finding_Rc}.  For example, the top vertex in panel (c) is vertex 3 and we denote it as $v_3$. We carry the same indexing over to the $\bvec{\psi}^\perp$-representation.

Ordering the items according to their distances from the simplex faces is easy in the $\bar{w}^\triangle$-representation: Because it provides barycentric coordinates, the distance of a point $i$ from the side opposite vertex $\alpha$ is just $w_\alpha(i)$, with the sign negative if the point lies outside the simplex. The $\bar{w}^\triangle$ ordering can be related to the $\bvec{\psi}^\perp$ ordering in a few steps. First, note that the distance of point $i$ from the side opposite $v_2$  is linearly related to the area of the triangle having vertices at points $i$, $v_1$, and $v_3$, with sign depending on triangle orientation. This signed area is proportional to the ratio of determinants
\[
A = \frac{|\bar{w}(i) \otimes \hat{\varepsilon}_1 \otimes \hat{\varepsilon}_3|}
{|\hat{\varepsilon}_2 \otimes \hat{\varepsilon}_1 \otimes \hat{\varepsilon}_3|} \;,
\]
where $\bar{w}(i) \otimes \hat{\varepsilon}_1 \otimes \hat{\varepsilon}_3$ is the $ 3 \times 3$ matrix formed by stacking the three row vectors and the denominator (which will always be $\pm 1$) ensures the correct sign. Second, note that since
\begin{eqnarray*}
 \bar{w}(i) \otimes \hat{\varepsilon}_1 \otimes \hat{\varepsilon}_3  & = & M \circ [\bvec{\psi}(i) \otimes \bvec{\psi}_{v_1} \otimes \bvec{\psi}_{v_3}]  \\
\protect \hat{\varepsilon}_2  \otimes \hat{\varepsilon}_1 \otimes \hat{\varepsilon}_3  & = & M \circ [\bvec{\psi}_{v_2} \otimes \bvec{\psi}_{v_1} \otimes \bvec{\psi}_{v_3}]  \;,
\end{eqnarray*}
where $\bvec{\psi}_{v_k}$ is the $m$-vector having the coordinates of vertex $v_k$ in the low-frequency eigenvector space,
\begin{equation}
\label{Eqn:A}
A = \frac{|\bvec{\psi}(i) \otimes \bvec{\psi}_{v_1} \otimes \bvec{\psi}_{v_3}|}
         {|\bvec{\psi}_{v_2} \otimes \bvec{\psi}_{v_1} \otimes \bvec{\psi}_{v_3}|} \;.
\end{equation}
Third, since all the $m$-vectors in \eqn{A} have their zeroth component equal to one, $A$ is proportional to the signed area of the triangle having vertices $i$, $v_1$, and $v_3$ in the $\bvec{\psi}^\perp$-representation. Fourth, this area is proportional to the distance of point $i$ from the side opposite to $v_2$ in the $\bvec{\psi}^\perp$-representation.  Combining all these proportionalities proves that the distance of point $i$ from the side opposing a vertex in the $\bar{w}^\triangle$-representation is proportional to its distance in the $\bvec{\psi}^\perp$-representation.
\section{Greedy algorithm for selecting ${\cal R}$}
\label{appendix:greedy}
The goal of the algorithm is to choose the subset of items $\cal R$ that approximately defines the $(m-1)$-simplex having maximum hypervolume $V_{m-1}$ in the $\bvec{\psi}^\perp$-representation. If the hypervolume, $V_{m-2}$, of one face of the simplex is already determined, $V_{m-1}$ is proportional to the distance of the excluded vertex from that face. [For example, in the case of a 2-simplex (a triangle), this is the familiar area $=\, 1/2$ base $\times$ height rule, where ``base'' is the length of the determined simplex face and ``height'' is the distance of the other point from that face.] This suggests a natural greedy algorithm: (a) initialize by finding the $(q-1=1)$-simplex of greatest length, (b) extend the $(q-1)$-simplex to a $q$-simplex by finding the item that is furthest from the hypersurface that embeds the $(q-1)$-simplex, (c) $q \leftarrow q+1$ and return to step (b) until $q=m$.

Specifically,
\begin{enumerate}
\item {\bf Initialize:}\\
 Select the two items $i_1$ and $i_2$ that  maximize
% Put this back for single-column/double-column
% \\
$||\bvec{\psi}^\perp(i_2) - \bvec{\psi}^\perp(i_1)||$.\\
                  ${\cal R} = \{i_1,i_2\}$.\\
                                    $q=2$.
\item {\bf Repeat while} $q < m${\bf:}\\
 (a) Select the item $i_{q+1}$ that maximizes \\
$ d^\perp(i_{q+1})  =  || {\cal P}^q \operp [\bvec{\psi}^\perp(i_{q+1})-\bvec{\psi}^\perp(i_1)] ||\,, $\\
\hspace*{1em} where \\
${\cal P}^q_{nn'}   =  I_{n n'} -$
% Put this back for single-column/double-column
% \\\hspace*{0.65in}
$\sum_{q'=2}^q \frac{[\bvec{\psi}^\perp(i_{q'})-\bvec{\psi}^\perp(i_1)]_n\, [\bvec{\psi}^\perp(i_{q'})-\bvec{\psi}^\perp(i_1)]_{n'} }
                                                          {|| \bvec{\psi}^\perp(i_{q'})-\bvec{\psi}^\perp(i_1) ||^2}\;. $\\
(b) ${\cal R} \leftarrow {\cal R} \cup i_{q+1}$ \\
(c) $q \leftarrow q+1$\\
\end{enumerate}

Here $\operp$ denotes the inner product within the $(m-1)$-dimensional $\bvec{\psi}^\perp$ space and ${\cal P}^q$ is the projection matrix in this space that removes the components of $[\bvec{\psi}^\perp(i_{q+1})-\bvec{\psi}^\perp(i_1)]$ that lie within the subspace containing the $(q-1)$-simplex. Therefore, $d^\perp(i_{q+1})$ is the distance of $\bvec{\psi}^\perp(i_{q+1})$ from the subspace, and the $q$-simplex formed by adding $\bvec{\psi}(i_{q+1})$ as a vertex is that of maximum hypervolume containing the previously computed $(q-1)$-simplex as one of its faces.

%\bibliography{../../../bibs/complete,footnote}

\begin{thebibliography}{76}
\bibitem[{ft:({\natexlab{a}})}]{ft:01}
\bibinfo{note}{Some authors analyze $I-\Gamma$, but this is an insignificant
  difference. Less frequently~\cite{Ng:02}, the ``normalized Laplacian''
  $\Gamma= D_\pi^{-1/2} \cdot S \cdot D_\pi^{-1/2}$ is analyzed, which does not
  satisfy \protect\eqns{diagonal_S} and
  \protect\eqnref{probability_conservation}. Such $\Gamma$ do not have a
  dynamical interpretation, so we do not consider them.}
\bibitem[{ft:({\natexlab{b}})}]{ft:02}
\bibinfo{note}{\protect\eqn{probability_conservation} is a direct consequence
  of \protect\eqn{diagonal_S}.}
\bibitem[{ft:({\natexlab{c}})}]{ft:03}
\bibinfo{note}{Although most spectral clustering $\Gamma$ satisfy
  \protect\eqns{dynamical_properties}, Pentney and Meila~\cite{Pentney:05} use
  a $\Gamma$ which does not and thereby gives rise to complex eigenvalues. We
  do not consider this uncommon situation.}
\bibitem[{ft:({\natexlab{d}})}]{ft:04}
\bibinfo{note}{Alternatively, the $\psi^R$ could be used, but the contrast
  between \protect\eqns{right_0_eigenvector} and \eqnref{left_0_eigenvector}
  makes it more convenient to use the $\psi^L$.}
\bibitem[{ft:({\natexlab{e}})}]{ft:05}
\bibinfo{note}{An exception is Ref.\ \cite{Alpert:99}, which uses more
  low-frequency eigenvectors than clusters.}
\bibitem[{ft:({\natexlab{f}})}]{ft:06}
\bibinfo{note}{In some cases, e.g., integrated circuit design, spectral methods
  have been used to partition items into a fixed number of subsets regardless
  of the existence of such a gap. We distinguish such forced partitioning from
  clustering and do not consider it here.}
\bibitem[{ft:({\natexlab{g}})}]{ft:07}
\bibinfo{note}{\protect\eqn{microscopic_dynamics} is the continuous master
  equation form used by Korenblum and Shalloway~\cite{Korenblum:03}. Other
  authors consider Markov processes, which are discretized versions of this
  equation. Our analysis applies with little change to the Markov case as
  well.}
\bibitem[{ft:({\natexlab{i}})}]{ft:09}
\bibinfo{note}{Ref.\ \protect\cite{Korenblum:03} actually uses the $D_\pi^{-1}
  \cdot \bfpsi_n^R$ as a basis, but by \protect\eqn{left_n_eigenvector} this is
  equivalent to using the $\bfpsi_n^L$.}
\bibitem[{ft:({\natexlab{j}})}]{ft:10}
\bibinfo{note}{Ref.\ \protect\cite{Korenblum:03} only explicitly considers the
  case when $(D_\pi)_{ii}=1/N$. Here we introduce the straightforward
  generalization that follows from their Eq.\ 10. Its inclusion in
  \protect\eqn{D_pi} ensures that items with little contribution to the
  equilibrium probability make little contribution to the uncertainty.}
\bibitem[{ft:({\natexlab{k}})}]{ft:11}
\bibinfo{note}{For $m>2$, the slave inequality constraint boundaries do not
  intersect at a point but in the subspace where
  \protect\eqn{coupled_inequality_constraints} is an equality for all items.
  This condition can be expressed as $\sum_{k=1}^{m(m-1)} Z_{ik} M^{\rm free}_k
  = 1 \;(\forall \; i)$, where $M_k^{\rm free} \equiv M_{\alpha n} \;
  \text{with} \; \alpha \equiv 1 + \Big \lfloor \frac{k-1}{m} \Big \rfloor$, $n
  \equiv (k-1) \bmod m$, and $Z_{ik} \equiv \psi_{k'}(i)$ with $k' \equiv (k-1)
  \bmod m$. ($M_k^{\rm free}$ is the flattened form of the supervector
  $\bvec{M}^{\rm free}$.) $Z_{ik}$ has $m$ independent columns, i.e., it
  contains $m-1$ copies of each of the $m$ low-frequency $\bfpsi_n$. Therefore,
  its rank is $m$ and it only imposes $m$ restrictions on the $m(m-1)$
  variables. Thus, the intersection of the inhomogeneous constraints occurs in
  a subspace of dimension $m(m-1)-m = m(m-2)$.}
\bibitem[{ft:({\natexlab{m}})}]{ft:13}
\bibinfo{note}{To prove this, we post-multiply \protect\eqn{M0_def} by
  $\Psi^{{\cal R}_c}$ and sum to get $1 = \sum_\alpha I_{\alpha \beta} =
  \sum_{\alpha n} M^0_{\alpha n} \Psi^{{\cal R}_c}_{n \beta} \; (\forall\;
  \beta)$. Since $\Psi^{\cal R}_{0 \beta} \equiv \psi_0(r_\beta) = 1 \;
  (\forall \, \beta)$, this is solved by $\sum_\alpha M^0_{\alpha n} =
  \delta_{n0}$. Moreover, since $\Psi^{\cal R}$ is invertible, this solution is
  unique, thereby proving \protect\eqn{M0_equality_constraint}.}
\bibitem[{ft:({\natexlab{n}})}]{ft:14}
\bibinfo{note}{Although they are constructed from the same components, do not
  confuse the $m$ $N$-vectors $\bfw^*_\alpha$, which are vectors over the space
  of items, with the $N$ $m$-vectors $\bar{w}^*(i)$, which are vectors over the
  space of clusters. (These vectors can be viewed as the rows and columns,
  respectively, of an $m \times N$ matrix $W^*$.)}.
\bibitem[{ft:({\natexlab{o}})}]{ft:15}
\bibinfo{note}{When $\Gamma$ is symmetric, the covariance matrix, $\Sigma_{mn}
  \equiv \bfpsi_m \cdot \bfpsi_n = \delta_{mn}$, is diagonal, so that the item
  points in the $\bvec{\psi}^\perp$-representation have principal components
  $\bfpsi_n \; (0 < n < m)$ and center of mass at the origin.}
\bibitem[{ft:({\natexlab{p}})}]{ft:16}
\bibinfo{note}{We ignore the rare case where LP has a space of degenerate
  solutions corresponding to a face of the feasible polytope.}
\bibitem[{ft:({\natexlab{r}})}]{ft:18}
\bibinfo{note}{In the spiral problem the representatives at vertices 1 and 2
  each contribute two active constraints to the solution $M^*$. Because the
  left side pivots out during refinement, the representative at vertex 3 only
  contributes one of the two constraints that are active at $M^0$; the second
  constraint associated with this vertex comes from the item identified by the
  arrow in \protect\fig{spiral_2d_initial_eig_rep}. Specifically, the
  constraints active at $M^*$ are $w_\alpha(i) = 0$ where $(\alpha, i) =
  (2,r_1), \, (3,r_1), \, (1,r_2), \, (3, r_2),\, (1,r_3),$ and $(2,x_3)$,
  where $x_3$ is the index of the item identified by the arrow. In this
  problem, these are all identified in a single step of refinement.}
\bibitem[{ft:({\natexlab{s}})}]{ft:19}
\bibinfo{note}{The number of eigenvectors computed is arbitrary and is chosen
  to limit computational cost on the assumption that the number of potential
  clusters to be identified at a single level of recursion is $< 20$. A larger
  value could be used if needed.}
\bibitem[{ft:({\natexlab{t}})}]{ft:20}
\bibinfo{note}{Ref.\ \cite{Weber:04} uses a slightly different variant in which
  the greedy algorithm is initialized by selecting the item that is farthest
  from the origin rather than the two items that are farthest from each other.
  This effectively maximizes the hypervolume of the $m$-simplex in the
  $\bvec{\psi}$ space, which includes the origin as one of the $m+1$ vertices,
  as opposed to maximizing the hypervolume of the $m-1$-simplex in the
  $\bvec{\psi}^\perp$-representation. Given that $\bfpsi_0 = \bfl$, all
  $m-1$-simplices in the $\bvec{\psi}^\perp$-representation lie in a plane
  equidistant from the origin. Hence, exact maximization in either space is
  equivalent. However, since the greedy algorithms may only approximately
  maximize the volumes, the two variants may give slightly different results.
  But such differences become irrelevant when uncertainty minimization is used
  because the greedy solution only provides the starting point for subsequent
  refinement.}
\bibitem[{ft:({\natexlab{u}})}]{ft:21}
\bibinfo{note}{The hypervolume in the $\bar{w}^\triangle$-representation of the
  $(m-1)$-simplex formed by any subset ${\cal R}$ of $m$ items (not necessarily
  representatives) is proportional to $|W^{\cal R}|$, which, according to
  \protect\eqn{w_mtrue_psi}, is \( |W^{\cal R}| = |M| \, |\Psi^{\cal R}| \,. \)
  Thus, the hypervolume for every subset is proportional to $|M|$, although the
  constant of proportionality, $|\Psi^{\cal R}|$, is different in each case.
  Therefore, maximizing $|M|$ maximizes the hypervolume of each
  $(m-1)$-simplex. Moreover, as discussed in Appendix
  \ref{appendix:bounding_triangle}, $\Psi^{\rm vert} = M^{-1}$, and since
  $|\Psi^{\rm vert}|$ is proportional to the area of the inequality constraint
  boundary $(m-1)$-simplex in the $\bvec{\psi}^\perp$-representation,
  maximizing $|M|$ minimizes the hypervolume of this simplex. To see that
  maximizing $|M|$ tends to equalize the cluster sizes, note that $|M| = |M
  \circ M^T|^{1/2} = |\Omega|^{1/2}$, where $\Omega_{\alpha \beta} \equiv
  \langle \protect\bfw_\alpha | \protect\bfw_\beta \rangle$. The off-diagonal
  elements of $\Omega$ are the equilibrium probability-weighted cluster
  overlaps and will be small compared to the diagonal elements. Therefore the
  determinant is dominated by the product of the diagonal elements,
  $\prod_\alpha \langle \protect\bfw_\alpha | \protect\bfw_\alpha \rangle
  = \prod_\alpha \langle (\protect\bfone - \sum_{\beta \ne \alpha}
  \protect\bfw_\beta) | \protect\bfw_\alpha \rangle \approx \prod_\alpha
  \langle \protect\bfone | \protect\bfw_\alpha \rangle$. This will be maximized
  when all the $\langle \protect\bfone | \protect\bfw_\alpha \rangle$ (i.e.,
  the weighted fractions of items assigned to each cluster) are equal.}
\bibitem[{ft:({\natexlab{v}})}]{ft:22}
\bibinfo{note}{Of course, $\gamma_{N-1}$ and $\gamma_1$ will be affected if
  extremal-rate processes occur in series or parallel with other processes of
  similar extremal magnitude. Thus, these are only rough guidelines, but they
  are sufficient for preconditioning, subject to the subsequent verification
  described in Sec.\ \protect\ref{section:truncating}.}
\end{thebibliography}
%\bibliographystyle{apsrev}

\end{document}